\documentclass[11pt,a4paper]{article}
\usepackage{amsmath,amsfonts,amssymb,amsthm}
\usepackage{fullpage,feynmp}

\newtheorem*{thm-ward}{Theorem \ref{thm:ward}}
\newtheorem{thm}{Theorem}

\newtheorem{lma}[thm]{Lemma} 
\newtheorem{prop}[thm]{Proposition}
\newtheorem{defn}[thm]{Definition}

\newtheorem{rem}[thm]{Remark}

\newcommand{\cross}[1]{~\#_#1~}
\def\Aut{\mathrm{Aut}}
\def\bar{\overline}

\def\card{\mathrm{card}}

\newcommand{\ins}[2]{#1~|~#2}
\def\e{{[1]}}
\def\eff{\mathrm{eff}}

\def\exter{\mathrm{ext}}

\def\Hom{\mathrm{Hom}}
\def\half{\tfrac{1}{2}}
\def\I{{\rm I}}
\def\II{{\rm II}}
\def\III{{\rm III}}
\def\id{\mathrm{id}} 

\def\inter{\mathrm{int}}
\def\isom{\simeq}

\def\lin{\mathrm{lin}}
 
\def\nn{\nonumber}

\def\res{\mathrm{res}}

\def\ST{\mathrm{ST}}
\def\Sym{\mathrm{Sym}}

\def\tilde{\widetilde}
\def\v{{[0]}}
\def\val{\mathrm{val}}

\def\el{
{~
    \begin{fmfgraph}(8,5)
      \fmfleft{l}
      \fmfright{r}
      \fmf{plain}{l,r}
    \end{fmfgraph}
  }
}

\def\ph{
{~
    \begin{fmfgraph}(8,5)
      \fmfleft{l}
      \fmfright{r}
      \fmf{photon}{l,r}
    \end{fmfgraph}
  }
}

\def\vertex{
{~
	\begin{fmfgraph}(8,5)
	  \fmfset{wiggly_len}{4pt} 
	  \fmfset{wiggly_slope}{70}
	  \fmfleft{l}
	  \fmfright{r1,r2}
	  \fmf{photon}{l,v}
	  \fmf{plain}{r1,v}
	  \fmf{plain}{v,r2}
	\end{fmfgraph}
      }
}

\def\qua{
{~
      \begin{fmfgraph}(8,6)
	\fmfpen{.1mm}
        \fmfforce{(0w,.15h)}{l}
        \fmfforce{(1w,.15h)}{r}
	\fmf{plain}{l,r}
      \end{fmfgraph}
}}

\def\gho{
{~
      \begin{fmfgraph}(8,6)
	\fmfpen{.1mm}
	\fmfset{dot_len}{.3mm}
        \fmfforce{(0w,.15h)}{l}
        \fmfforce{(1w,.15h)}{r}
        	\fmf{dots}{l,r}
    \end{fmfgraph}
}}

\def\glu{
{~
      \begin{fmfchar}(8,6)
	\fmfpen{.1mm}
	\fmfset{curly_len}{.5mm}
	\fmfleft{l}
	\fmfright{r}
	\fmf{gluon}{l,r}
      \end{fmfchar}
}}

\def\quaglu{
{~
      \begin{fmfchar}(8,6)
	\fmfpen{.1mm}
	\fmfset{curly_len}{.5mm}
	\fmfleft{l}
	\fmfright{r1,r2}
	\fmf{gluon}{l,v}
	\fmf{plain}{r1,v}
	\fmf{plain}{v,r2}
      \end{fmfchar}
}}

\def\ghoglu{
{~
      \begin{fmfchar}(8,6)
	\fmfpen{.1mm}
	\fmfset{curly_len}{.5mm}
	\fmfset{dot_len}{.3mm}
	\fmfleft{l}
	\fmfright{r1,r2}
	\fmf{gluon}{l,v}
	\fmf{dots}{r1,v}
	\fmf{dots}{v,r2}
      \end{fmfchar}
}}

\def\gluc{
{~
      \begin{fmfchar}(8,6)
	\fmfpen{.1mm}
	\fmfset{curly_len}{.5mm}
	\fmfleft{l}
	\fmfright{r1,r2}
	\fmf{gluon}{l,v}
	\fmf{gluon}{v,r1}
	\fmf{gluon}{v,r2}
      \end{fmfchar}
}}

\def\gluq{
{~
      \begin{fmfchar}(8,6)
	\fmfpen{.1mm}
	\fmfset{curly_len}{.5mm}
	\fmfleft{l1,l2}
	\fmfright{r1,r2}
	\fmf{gluon}{l1,v}
	\fmf{gluon}{l2,v}
	\fmf{gluon}{r1,v}
	\fmf{gluon}{v,r2}
      \end{fmfchar}
}}

\title{Renormalization of gauge fields: \\A Hopf algebra approach}
\author{
Walter D. van Suijlekom\\[7mm]
Max Planck Institute for Mathematics\\
Vivatsgasse 7, D-53111 Bonn, Germany\\
\texttt{waltervs@mpim-bonn.mpg.de}
}
\date{Oktober 12, 2006}

\begin{document}
\begin{fmffile}{graphs-st}

\fmfset{wiggly_len}{5pt} 
\fmfset{wiggly_slope}{70} 
\fmfset{curly_len}{1.4mm}

\fmfset{dot_len}{1mm}

\maketitle

\begin{abstract}
We study the Connes--Kreimer Hopf algebra of renormalization in the case of gauge theories. We show that the Ward identities and the Slavnov--Taylor identities (in the abelian and non-abelian case respectively) are compatible with the Hopf algebra structure, in that they generate a Hopf ideal. Consequently, the quotient Hopf algebra is well-defined and has those identities built in. 
This provides a purely combinatorial and rigorous proof of compatibility of the Slavnov--Taylor identities with renormalization. 
\end{abstract}

\section{Introduction}
The combinatorial structure underlying renormalization in perturbative quantum field theory was transparent in the original approach of Bogogliubov, Hepp, Parasiuk and Zimmerman (cf. for instance \cite[Ch.5]{Col84}). It was realized by Kreimer in \cite{Kre98} that this structure is in fact organized by a Hopf algebra of rooted trees. One year later, Connes and Kreimer \cite{CK99} reformulated this combinatorial structure directly in terms of a (commutative) Hopf algebra of Feynman graphs and understood the BPHZ-procedure as a Birkhoff decomposition in the group that is dual to this Hopf algebra. 

In physics, however, one usually works in the setting of functional calculus since -- although defined only formally -- functional integrals are particularly well-suited for the perturbative treatment of quantum gauge theories. For example, the Slavnov--Taylor identities that are the reminiscents of the gauge symmetry of the classical field theory, can be gathered elegantly in a single equation known as the {\it Zinn-Justin equation}, and involving the effective action.
Although this approach is very powerful in showing for example renormalizability of gauge field theories, the applicability of the graph-per-graph approach of the BPHZ-renormalization procedure is not so transparent any more, and the same holds for the combinatorial structure underlying it. 

Recent developments \cite{Kre05,Kre06} give more insight in the combinatorial aspects of non-abelian gauge theories. In \cite{Sui06}, we considered the Hopf algebra of Feynman graphs in quantum electrodynamics and found that certain Ward--Takahashi identities can be imposed as relations on this Hopf algebra.

In this article, we continue to explore the combinatorial structure of gauge theories in terms of a Hopf algebra. More precisely, we will show that the Slavnov--Taylor identities between the coupling constants can be implemented as relations on the Hopf algebra, that is, in a way compatible with the counit, coproduct and antipode. This then provides a combinatorial proof of the compatibility of the Slavnov--Taylor identities with renormalization.

We start in Section \ref{sect:prel} with a precise setup of the Hopf algebra of Feynman graphs in a generic theory, including gauge theories, and derive a formula for the coproduct on 1PI Green's functions. Such Green's functions are sums of all 1PI graphs with a certain fixed external structure, including symmetry factors. It is the latter that make this derivation slightly involved. 

Section \ref{sect:ward} we will be a warming-up for the non-abelian case, by considering quantum electrodynamics, which is an abelian gauge theory. Using the expression for the coproduct on the Green's functions, we show that certain Ward identities can be imposed as (linear) relations on the Hopf algebra. In other words, they define a Hopf ideal. 

The case of non-abelian gauge theories will be consided in Section \ref{sect:st}, where we will show that the Slavnov--Taylor identities define quadratic relations in the Hopf algebra. In fact, as we will see, it is in the very nature of the combinatorial factors that are involved that the Slavnov--Taylor identities appear. 

We have added two appendices. In the first, we rederive the compatibility of Ward identities with the Hopf algebra structure in QED obtained in Section \ref{sect:ward} from our previous result on Ward--Takahashi identities in \cite{Sui06}. In the second appendix, we list some useful basic combinatorial identities used throughout the text.

\section{General structure of the Hopf algebra of Feynman graphs}
\label{sect:prel}
We start with some definitions on Feynman graphs and their symmetries, thereby making precise several properties needed later.
\subsection{Feynman graphs}
The Feynman graphs we will consider are built from a certain set of edges and vertices $R$, and we write $R=R_V \cup R_E$. For example, in $\phi^3$-theory, the set $R_V$ contains the bi- and trivalent vertex and $R_E$ the straight line, but more interesting theories such as gauge theories contain different types of edges and vertices (for example involving curly, dotted and straight lines) corresponding to different particles. More precisely, we have the following definition \cite{CM07}.
\begin{defn}
A {\rm Feynman graph} $\Gamma$ is given by a set $\Gamma^\v$ of vertices each of which is an element in $R_V$ and $\Gamma^\e$ of edges in $R_E$, and maps 
$$
\partial_j: \Gamma^\e \to \Gamma^\v \cup \{ 1, 2, \ldots, N\}, \qquad j=0,1,
$$
that are compatible with the type of vertex and edge as parametrized by $R_V$ and $R_E$, respectively. The set $\{ 1,2, \ldots, N\}$ labels the external lines, so that $\sum_j \card~ \partial_j^{-1} (v) =1$ for all $v \in \{1, \ldots,N\}$.

The set of {\rm external lines} is $\Gamma_\exter^\e = \cup_i \partial_i^{-1} \{1, \ldots, N\}$ and its complement $\Gamma_\inter^\e$ in $\Gamma^\e$ is the set of {\rm internal lines}. 
\end{defn}
We remark that the elements in $\Gamma^\e_\exter$ can thus be labeled as $e_1, \ldots, e_N$ where $e_k :=\cup_i \partial_i^{-1}(k)$ and we understand this labeling as being fixed.
With this definition, the notion of a graph automorphism can be defined as follows.
\begin{defn}
An {\rm automorphism of a Feynman graph $\Gamma$} is given by an isomorphism $g^\v$ from $\Gamma^\v$ to itself, and an isomorphism $g^\e$ from $\Gamma^\e$ to itself that is the identity on $\Gamma^\e_\exter$ and such that for all $e \in \Gamma^\e$,
\begin{equation}
\label{eq:graph-auto}
\cup_j g^\v (\partial_j(e))= \cup_j \partial_j(g^\e(e)).
\end{equation}
Moreover, we require $g^\v$ and $g^\v$ to respect the type of vertex/edge in the set $R$.\\[2mm]
The {\rm automorphism group} $\Aut(\Gamma)$ of $\Gamma$ consists of all such automorphisms; its order is called the {\rm symmetry factor} of $\Gamma$ and is denoted by $\Sym(\Gamma)$. 
\end{defn}
Similarly, there is a notion of an isomorphism of two graphs $\Gamma$ and $\Gamma'$ as a pair of maps that intertwines the maps $\partial_i$ as in Eq. \eqref{eq:graph-auto}. We remark that we correct in this way for the apparent orientation given by the two maps $\partial_0$ and $\partial_1$ and we stress that the fermionic lines are unoriented. We take the complex character of the fermionic fields into account by summing over all possible orientations once we apply the Feynman rules. 

The above definition of automorphism differs from the usual notion of graph automorphism (cf. for instance \cite{Die97}) in that the latter might also permute the elements in $\{1, \ldots,N\}$ when understood as external vertices. In the above notation, such an automorphism of $\Gamma$ would be given by an isomorphism $g^\v$ from $\Gamma^\v \cup \{1, \ldots,N\}$ to itself, and an isomorphism $g^\e$ from $\Gamma^\e$ to itself such that Equation \eqref{eq:graph-auto} holds.
\begin{figure}[t]
\begin{center}
$\Sym\big( $
  \parbox{40pt}{
\begin{fmfgraph*}(40,50)
  \fmfleft{l}
  \fmfright{r}
  \fmf{photon}{l,v,r}
  \fmfv{decor.shape=circle, decor.filled=0, decor.size=5thick}{v}
  \end{fmfgraph*}
}$~\big)=2$
\hspace{3cm}
$\Sym\big( $
\parbox{40pt}{
\begin{fmfgraph*}(40,50)
      \fmfleft{l}
      \fmfright{r}
      \fmf{plain}{l,v1,v2,r}
      \fmf{photon,left,tension=0}{v1,v2}
\end{fmfgraph*}
}$~\big)=1$
\end{center}
\caption{Automorphisms of Feynman graphs respect the type of vertex/edge in $R$. }
\end{figure}

\bigskip 

Note that for $\Gamma=\prod_i \Gamma_i$ the disjoint union of $n$ graphs, the symmetry factor is given by $\Sym(\Gamma) = n_1! \cdots n_k! ~\Sym(\Gamma_{1}) \cdots \Sym(\Gamma_{n})$ where $n_i$ are the numbers of isomorphic (with fixed external lines) connected components of $\Gamma'$. Equivalently, one has for a 1PI graph $\Gamma'$, 
\begin{equation}
\label{eq:sym-union}
\Sym(\Gamma ~\Gamma') =  n(\Gamma, \Gamma') \Sym(\Gamma) \Sym(\Gamma'),
\end{equation}
with $n(\Gamma, \Gamma')$ the number of connected components of $\Gamma \Gamma'$ that are isomorphic to $\Gamma'$.

\bigskip

If $\Gamma$ is a connected Feynman graph with external lines labeled by $\{1, \ldots, N\}$, we can construct another graph $\Gamma^\sigma$, by permutating the external lines by an element $\sigma$, respecting the type of external lines. The graph $\Gamma^\sigma$ is given by the same sets $\Gamma^\v$ and $\Gamma^\e$ but with maps
$$
\partial_j^\sigma := \sigma \circ \partial_j : \Gamma^\e \to \Gamma^\v \cup \{1, \cdots, N\}. 
$$
This permutation affects the labeling of the external lines by $\{1, \ldots N\}$, which explains the terminology permutation of external lines; we write $e^\sigma$ for the edge in $\Gamma^\sigma$ corresponding to an edge $e \in \Gamma^\e$ under the permutation $\sigma$.
\begin{defn}
\label{defn:triv-perm}
A permutation $\sigma$ of the external lines of $\Gamma$ is called {\rm trivial} if there exists an isomorphism between $\Gamma^\sigma$ and $\Gamma$, leaving the labeling of the external lines fixed.\\ 
The number of non-isomorphic graphs $\Gamma^\sigma$ obtained by a permutation $\sigma$ of the external lines of $\Gamma$, is denoted by $|\Gamma|_\vee$ and extended to disconnected graphs by $|\Gamma \Gamma'|_\vee= |\Gamma|_\vee |\Gamma'|_\vee$. 
\end{defn}
\begin{lma}
\label{lma:perm}
A permutation $\sigma$ of the external lines of $\Gamma$  is trivial if and only if there exists an automorphism $g$ of the graph $\Gamma$ not necessarily leaving the external lines fixed, such that $g^\v|_{\{1,\ldots,N\}}=\sigma$.
\end{lma}
\begin{proof}
Firstly, if $\sigma$ is trivial, there exists an isomorphism $f: \Gamma^\sigma \to \Gamma$ and the pair $(f^\v \circ \sigma,~f^\e \circ \sigma)$ is an automorphism $g$ of $\Gamma$ (without fixed external vertices), since,
\begin{align*}
\cup_j g^\v (\partial_j(e)) = \cup_j f^\v (\partial^\sigma_j(e^\sigma)) =\cup_j \partial_j(f^\e(e^\sigma)) = \cup_j \partial (g^\e(e)). 
\end{align*}
On the other hand, such an automorphism $g$ is given by two maps $g^\v$ and $g^\e$, where $g^\v$ is the product of two permutations of the disjoint sets $\Gamma^\v$ and $\{1, \ldots,N\}$, say $f^\v$ and $\sigma$, respectively. Correspondingly, $\sigma$ acts on $\Gamma^\e_\exter$ by permutation, so that also $g^\e=f^\e \circ \sigma$. This factorization gives rise to an isomorphism $f$ from $\Gamma^\sigma$ to $\Gamma$, which leaves external lines fixed. 
\end{proof}
\begin{figure}[h!]
\begin{center}
\parbox{60pt}{
  \begin{fmfgraph*}(60,60)
      \fmfleft{l}
      \fmfright{dr1,r1,r,r2,dr2}
      \fmflabel{1}{l}
      \fmflabel{3}{r1}
      \fmflabel{2}{r2}
      \fmf{photon}{l,v}
      \fmf{phantom}{v,v1,dr1}
      \fmf{phantom}{v,v2,dr2}
      \fmffreeze
      \fmf{plain}{v,v1,r1}
      \fmf{plain}{v,v2,r2}
      \fmffreeze
      \fmf{photon}{v1,loop,v2}
      \fmffreeze
      \fmfv{decor.shape=circle, decor.filled=0, decor.size=4thick}{loop}
      \fmf{dots}{v,r}
   \end{fmfgraph*}
}
\hspace{2cm}
$\overset{\sigma}{\longrightarrow}$
\hspace{2cm}
\parbox{60pt}{
 \begin{fmfgraph*}(60,60)
      \fmfleft{l}
      \fmfright{dr1,r1,r,r2,dr2}
      \fmflabel{1}{l}
      \fmflabel{3}{r1}
      \fmflabel{2}{r2}
      \fmf{photon}{l,v}
      \fmf{phantom}{v,v1,dr1}
      \fmf{phantom}{v,v2,dr2}
      \fmffreeze
      \fmf{plain}{v,v1,r2}
      \fmf{plain}{v,v2,r1}
      \fmf{photon}{v1,loop,v2}
      \fmffreeze
      \fmfv{decor.shape=circle, decor.filled=0, decor.size=4thick}{loop}
    \end{fmfgraph*}
}
\caption{The permuation $\sigma=(23)$ of the external lines of the graph $\Gamma$ is trivial since reflection in the dotted line induces an automorphism $g$ of $\Gamma$ such that $g^\v|_{\{1,2,3\}} = \sigma$. Moreover, this is the only trivial permutation so that $|\Gamma|_\vee= 3!/2=3$}
\end{center}
\end{figure}
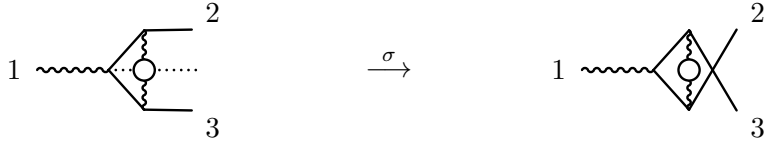
\bigskip

For the purpose of renormalization, one is mainly interested in one-particle irreducible Feynman graphs which have {\it residues} that are elements in the set $R$. 
\begin{defn}
A Feynman graph is called {\rm one-particle irreducible (1PI)} if it is not a tree and can not be disconnected by removal of a single edge.
\end{defn}
\begin{defn}
The {\rm residue} $\res(\Gamma)$ of a Feynman graph $\Gamma$ is defined as the vertex/edge the graph corresponds to after collapsing all its internal edges to a point. 
\end{defn}
\noindent
For example, we have
\begin{align*}
\res\left( \parbox{40pt}{
\begin{fmfgraph*}(40,30)
      \fmfleft{l}
      \fmfright{r1,r2}
      \fmf{photon}{l,v}
      \fmf{plain}{v,v1,r1}
      \fmf{plain}{v,v2,r2}
      \fmffreeze
      \fmf{photon}{v1,loop,v2}
      \fmffreeze
      \fmfv{decor.shape=circle, decor.filled=0, decor.size=2thick}{loop}
\end{fmfgraph*}
}\right) = 
\parbox{20pt}{
\begin{fmfgraph*}(20,20)
      \fmfleft{l}
      \fmfright{r1,r2}
      \fmf{photon}{l,v}
      \fmf{plain}{v,r1}
      \fmf{plain}{v,r2}
\end{fmfgraph*}
}
\qquad \text{ and }\qquad
\res\left( \parbox{40pt}{
\begin{fmfgraph*}(40,30)
      \fmfleft{l}
      \fmfright{r}
      \fmf{plain}{l,v1,v2,v5,v6,r}
      \fmf{photon,left,tension=0}{v1,v5}
      \fmf{photon,right,tension=0}{v2,v6}
\end{fmfgraph*}
}\right) = 
\parbox{20pt}{
\begin{fmfgraph*}(20,20)
      \fmfleft{l}
      \fmfright{r}
      \fmf{plain}{l,r}
\end{fmfgraph*}
}~.
\end{align*}
We restrict to the class of Feynman graphs $\Gamma$ for which $\res(\Gamma) \in R$ and will denote a generic graph with residue $r \in R$ by $\Gamma^r$. If it also has loop number $L$, we denote it by $\Gamma^r_L$.
\begin{defn}
The Hopf algebra $H$ of Feynman graphs is the free commutative algebra generated by all 1PI Feynman graphs, with counit $\epsilon(\Gamma)=0$ unless $\Gamma=\emptyset$, in which case $\epsilon(\emptyset)=1$, coproduct,
\begin{align}
\label{coproduct}
\Delta (\Gamma) = \Gamma \otimes 1 + 1 \otimes \Gamma + \sum_{\gamma \subsetneq \Gamma} \gamma \otimes \Gamma/\gamma,
\end{align}
and antipode given recursively by,
\begin{equation}
\label{antipode}
S(\Gamma) = - \Gamma - \sum_{\gamma \subsetneq \Gamma} S(\gamma) \Gamma/\gamma.
\end{equation}
\end{defn}

\subsection{Insertion of graphs}
\label{sect:ins}
\begin{defn}
\label{defn:ins}
An {\rm insertion place} for a (connected) graph $\gamma$ in $\Gamma$ is the subset of $\Gamma^{[0]} \cup \Gamma^{[1]}$ consisting of vertices/internal edges of the form $r = \res(\gamma)$. It can be extended to disconnected graphs $\gamma=\prod_{i=}^n \gamma_i$ by giving $n$-tuples of insertion places for $\gamma_1, \ldots, \gamma_n$, thereby allowing several insertions of the connected components with residue $r$ in $R_E$ on the same internal edge in $\Gamma$ of the form $r$. The number of insertion places for $\gamma$ in $\Gamma$ is denoted by $\ins{\Gamma}{\gamma}$.
\end{defn}
An explicit expression for $\ins{\Gamma}{\gamma}$ can be obtained as follows \cite{Kre05}. Let $m_{\Gamma, r}$ be the number of vertices/edges $r$ in $\Gamma^{[0]} \cup \Gamma^{[1]}$, for $r \in R$. Moreover, let $n_{\gamma,r}$ be the number of connected components of $\gamma$ with residue $r$. Since insertion of a vertex graph ({i.e.} with residue in $R_V$) on a $v \in \Gamma^{[0]}$ prevents a subsequent insertion at $v$ of a vertex graph with the same residue, whereas insertion of an edge graph ({i.e.} with residue in $R_E$) creates two new egdes and hence two insertion places for a subsequent edge graph, we find the following expression,
\begin{align*}
\ins{\Gamma}{\gamma} 
&=\prod_{v \in R_V} n_{\gamma,v}! {m_{\Gamma,v}  \choose n_{\gamma,v}} \prod_{e \in R_E} n_{\gamma,e}! {m_{\Gamma,e}+n_{\gamma,e}-1  \choose n_{\gamma,e}}.
\end{align*}
Indeed, the binomial coefficients arises for each vertex $v$ since we are choosing $n_{\gamma,v}$ out of $m_{\Gamma,v}$ whereas for an edge $e$ we choose $n_{\gamma,e}$ out of $m_{\Gamma,e}$ {\it with repetition}.
We extend this definition to empty graphs by defining $\ins{\Gamma}{\emptyset} =  \ins{\emptyset}{\gamma}= \ins{\emptyset}{\emptyset}=1$ for a 1PI graph $\gamma$, and $\ins{\emptyset}{\gamma} = 0$ for a disconnected graph $\gamma$. 
\begin{rem}
Our expression for $\ins{\Gamma}{\gamma}$ differs slightly from the one given in \cite{Kre05} where additional factors of $1/n_{\gamma,r}!$ are present for $r \in R$. 
It turns out that the above expression appears naturally in the coproduct on 1PI Green's functions (see below).  
\end{rem}
\noindent
A few examples are in place:
\begin{align*}
\parbox{30pt}{
  \begin{fmfgraph*}(30,40)
  \fmfleft{l}
  \fmfright{r}
  \fmf{photon}{l,v,r}
  \fmfv{decor.shape=circle, decor.filled=0, decor.size=5thick}{v}
  \end{fmfgraph*}
} 
~\Big|~
\parbox{40pt}{
\begin{fmfgraph*}(40,40)
      \fmfleft{l}
      \fmfright{r1,r2}
      \fmf{photon}{l,v}
      \fmf{plain}{v,v1,r1}
      \fmf{plain}{v,v2,r2}
      \fmffreeze
      \fmf{photon}{v1,v2}
\end{fmfgraph*}
} ~ = {2 \choose 1} = 2
\quad \text{ whereas }\quad 
\parbox{30pt}{
  \begin{fmfgraph*}(30,40)
  \fmfleft{l}
  \fmfright{r}
  \fmf{photon}{l,v,r}
  \fmfv{decor.shape=circle, decor.filled=0, decor.size=5thick}{v}
  \end{fmfgraph*}
} 
~\Big|~
\parbox{30pt}{
\begin{fmfgraph*}(30,30)
      \fmfleft{l}
      \fmfright{r}
      \fmf{plain}{l,v1,v2,r}
      \fmf{photon,left,tension=0}{v1,v2}
\end{fmfgraph*}
} ~ 
\parbox{30pt}{
\begin{fmfgraph*}(30,30)
      \fmfleft{l}
      \fmfright{r}
      \fmf{plain}{l,v1,v2,v5,v6,r}
      \fmf{photon,left,tension=0}{v1,v5}
      \fmf{photon,right,tension=0}{v2,v6}
\end{fmfgraph*}
} ~ = 2! {3 \choose 2}  = 6.
\end{align*}
\begin{defn}
An {\rm insertion} of a connected graph $\gamma$ at the insertion place $x$ in $\Gamma$, is given by a bijection between the set $\gamma^\e_\exter$ of external lines of $\gamma$ and the set $\partial^{-1}(x)$. If $x \in \Gamma^\v$, $\partial^{-1}(x)$ denotes the set of lines attached to the vertex $x$, and if $x \in \Gamma^\e_\exter$, $\partial^{-1}(x)$ denotes the set of adjacent edges to any internal point of $x$. The graph obtained in this way is denoted by $\Gamma \circ_{(x,\phi)} \gamma$.\\[2mm]
Two insertions $(x,\phi)$ and $ (x',\phi')$ are called equivalent if $x=x'$ and $\phi'=\phi\circ \sigma$ for some trivial permutation $\sigma$ of the external lines of $\gamma$. The set of all insertions of $\gamma$ in $\Gamma$ up to equivalence is denoted by $X(\Gamma,\gamma)$; it consists of equivalence classes $[x,\phi]$. 
\end{defn}
This equivalence relation on insertions makes sense in that $\Gamma \circ_{(x,\phi)} \gamma \isom \Gamma \circ_{(x',\phi')} \gamma$ whenever $(x,\phi) \sim (x',\phi')$.
We extend $X(\Gamma, \gamma)$ to disconnected graphs $\gamma$ as follows. If $\gamma= \prod_{i=1}^n \gamma_i$ is the disjoint union of $n$ graphs, the set $X(\Gamma, \gamma)$ of insertions of $\gamma$ in $\Gamma$ is defined as the set of $n-$tuples of pairs $\left( [x_1,\phi_1], \ldots, [x_n,\phi_n]\right)$, where $[x_1,\phi_1] \in X(\Gamma, \gamma_1)$ and $[x_{k+1},\phi_{k+1}]$ is an element in $X(\Gamma \circ_{(x_1,\phi_1)\ldots(x_k,\phi_k)} \prod_{i=1}^k \gamma_i, \gamma_{k+1})$ which is not part of any of the inserted graphs $\gamma_1,\ldots,\gamma_{k-1}$ for $k=1,\ldots,n-1$. 
The cardinality of $X(\Gamma, \gamma)$ is the number $\ins{\Gamma}{\gamma}$ of insertion places for $\gamma$ in $\Gamma$ times the number $|\gamma|_\vee$ of non-trivial permutations of the external lines of $\gamma$.

We also need the following generalization for the number of insertion places.
\begin{defn}
\label{defn:ins3}
Let $\Gamma, \gamma, \gamma'$ be three (disjoint unions of) 1PI graphs. We define $\ins{\Gamma}{\ins{\gamma}{\gamma'}}$ to be the number of places to insert $\gamma$ into $\Gamma$ (say, at $x$ using $\phi$) and then subsequently insert $\gamma'$ in $\Gamma \circ_{(x,\phi)} \gamma$. In other words,
$$
\ins{\Gamma}{\ins{\gamma}{\gamma'}} := \frac{1}{|\gamma|_\vee} \sum_{[x,\phi] \in X(\Gamma,\gamma)}\ins{\Gamma \circ_{(x,\phi)} \gamma}{\gamma'}.
$$
Moreover, we set $\ins{\Gamma}{\ins{\emptyset}{\gamma'}} = \ins{\Gamma}{\gamma'}$ and $\ins{\emptyset}{\ins{\gamma}{\gamma'}} =1$ for two 1PI graphs $\gamma,\gamma'$ and $\ins{\emptyset}{\ins{\gamma}{\gamma'}}=0$ if one of the two graphs $\gamma,\gamma'$ is disconnected.
\end{defn}
The factor $1/|\gamma|_\vee$ corrects for the overcounting due to the several (non-equivalent) ways to insert $\gamma$ into $\Gamma$ at a particular place.
Note that automatically $\ins{\Gamma}{\ins{\gamma}{\emptyset}} = \ins{\Gamma}{\gamma}$ and if $\Gamma,\gamma,\gamma' \neq \emptyset$, we have
\begin{equation}
\label{eq:ins3-ins}
\ins{\Gamma}{\ins{\gamma}{\gamma'}} = \ins{\Gamma}{\gamma \gamma'} + (\ins{\Gamma}{\gamma})(\ins{\gamma}{\gamma'}).
\end{equation}

\bigskip

Suppose $\gamma$ is a 1PI graph. There is a natural action of $\Aut(\Gamma)$ on $X(\Gamma, \gamma)$ given by
\begin{equation*}
g \cdot (x,\phi) = (g x, g \circ \phi).
\end{equation*}
One readily checks that this action respects the equivalence relation on insertions, and therefore acts on the equivalence classes $[x,\phi]$. Moreover, an element $g \in \Aut(\Gamma)$ naturally induces an isomorphism $\Gamma \circ_{(x,\phi)} \gamma$ to $\Gamma \circ_{g(x,\phi)} \gamma$
For an element $[x,\phi]$ in $X(\Gamma, \gamma)$, we denote by $M(x,\phi)$ the number of graphs $\gamma'$ in $\Gamma \circ_{(\phi,x)} \gamma$ that are images of $\gamma$ under some element in $\Aut(\Gamma \circ_{(x,\phi)} \gamma)$. Moreover, $N(x,\phi)$ denotes the number of orbits $\Aut(\Gamma)[x',\phi']$ such that $\Gamma \circ_{(x',\phi')} \gamma \isom \Gamma \circ_{(x,\phi)} \gamma $. Both definitions are independent of the choice of a representative $(x,\phi)$ as well as the choice of the element $[x',\phi']$ in the orbit. Indeed, an element $g$ in $\Aut(\Gamma)$ will induce a natural isomorphism $ \Gamma \circ_{(x',\phi)} \gamma \isom \Gamma \circ_{g(x',\phi')} \gamma$.

\begin{lma}
\label{lma:orbit}
Suppose $\gamma$ is a 1PI graph and let $x \in X(\Gamma,\gamma)$. 
The length of the orbit $\Aut(\Gamma)[x,\phi]$ is given by
\begin{equation*}
\left| \Aut(\Gamma) [x,\phi] \right| = \frac{\Sym(\gamma) \Sym(\Gamma) M(x,\phi)}{\Sym(\Gamma \circ_{(x,\phi)} \gamma)}.
\end{equation*}
\end{lma}
\begin{proof}
We use the orbit-stabilizer theorem, stating in this case that the orbit $\Aut(\Gamma)[x,\phi]$ is isomorphic to the left cosets of the stabilizer $\Aut( \Gamma)_{[x,\phi]}$ in $\Aut(\Gamma)$. In particular, we have for its length,
\begin{equation*}
|\Aut(\Gamma)[x,\phi]|= \left[\Aut( \Gamma): \Aut( \Gamma)_{[x,\phi]} \right]=\frac{|\Aut(\Gamma)|}{|\Aut(\Gamma)_{[x,\phi]}|}.
\end{equation*}
The order of $\Aut( \Gamma)_{[x,\phi]}$ can be computed as follows. Let $\Aut(\Gamma\circ_{(x,\phi)} \gamma)_\gamma$ be the subgroup of $\Aut(\Gamma \circ_{(x,\phi)} \gamma)$ consisting of automorphisms that map $\gamma$ to itself (but possibly permuting the external lines of $\gamma$). There is a short exact sequence of groups
\begin{equation*}
1 \to \Aut(\gamma) \to \Aut(\Gamma\circ_{(x,\phi)} \gamma)_\gamma \to \Aut( \Gamma)_{[x,\phi]} \to 1.
\end{equation*}
Indeed, the image $\tilde g$ inside $\Aut(\Gamma)$ of an element $g$ in $\Aut(\Gamma \circ_{(x,\phi)} \gamma)_\gamma$ is defined by restricting $g$ to $\Gamma-\{x\}$ and by the identity map on the vertex $x$. Then, by Lemma \ref{lma:perm}, $\tilde g$ might permute the edges connected to the vertex $x$ but always in a trivial way, since $g$ induces an automorphism of $\gamma$ not necessarily leaving its external lines fixed. Therefore, $\tilde g(x, \phi) = (x,\phi \circ \sigma)$ for some trivial permutation  $\sigma$ of $\gamma_\exter^\e$, so that it is an element in the fixed point subgroup $\Aut(\Gamma)_{[x,\phi]}$. Moreover, the kernel of the map that sends such a $g$ to $\tilde g$ consists precisely of those elements in $\Aut(\Gamma \circ_{(x,\phi)} \gamma)_\gamma$ that correspond to the identity on $\Gamma$; in other words, these are automorphisms of $\gamma$ that leave external lines fixed.

We conclude that the quotient group $\Aut(\Gamma\circ_{(x,\phi)} \gamma)_\gamma / \Aut(\gamma)$ is isomorphic to $\Aut(\Gamma)_{[x,\phi]}$. Since $\Aut(\Gamma\circ_{(x,\phi)} \gamma)$ is generated by the elements in $\Aut(\Gamma\circ_{(x,\phi)} \gamma)_\gamma$ and automorphisms that map $\gamma$ isomorphically to a subgraph $\gamma'$ of $\Gamma$, we see that 
\begin{equation*} 
| \Aut(\Gamma \circ_{(x,\phi)} \gamma)_\gamma | = \frac{|\Aut(\Gamma \circ_{(x,\phi)} \gamma)|}{M(x,\phi)}.
\end{equation*}
Combining these results, we conclude that
\begin{equation*}
|\Aut(\Gamma)[x,\phi]|= \frac{|\Aut(\gamma)|~|\Aut( \Gamma)|}{|\Aut(\Gamma\circ_{(x,\phi)}\gamma)_\gamma|} = \frac{\Sym(\gamma)\Sym(\Gamma)M(x,\phi)}{\Sym(\Gamma\circ_{(x,\phi)} \gamma)}.
\end{equation*}
\end{proof}
As a final preparation to the next section, we will write the coproduct as a sum of maps $\Delta_\gamma$, with $\gamma$ a disjoint union of 1PI graphs (with fixed external lines). It is given by 
\begin{equation}
\Delta_\gamma(\Gamma) = \sum_{\gamma' \subset \Gamma, \gamma' \isom \gamma } \Gamma/\gamma',
\end{equation}
and defined to be zero if $\Gamma$ contains no subgraphs isomorphic to $\gamma$. 
In particular, $\Delta_\emptyset$ is the identity map, $\Delta_\Gamma(\Gamma)=\emptyset$ and $\Delta_\gamma(\emptyset)=0$ if $\gamma \neq \emptyset$. However, since only subgraphs isomorphic to $\gamma$ enter in this formula -- hence no reference is made to a particular labeling of the external lines of $\gamma$ -- we have to correct by a factor of $|\gamma|_\vee$ if we are to sum over all disjoint unions of 1PI graphs {\it with fixed external lines},
$$
\Delta= \sum_\gamma \frac{1}{|\gamma|_\vee} ~  \gamma \otimes \Delta_\gamma.
$$ 
We recall the following combinatorial factor from \cite{CK99}; for a given $\Gamma, \gamma, \Gamma'$, we denote by $n(\Gamma,\gamma,\Gamma')$ the number of subgraphs $\gamma' \isom \gamma$ in $\Gamma$ such that $\Gamma/\gamma \isom \Gamma'$. With this definition, we can write
\begin{equation}
\label{eq:cop-gamma-comb}
\Delta_\gamma(\Gamma) = \sum_{\Gamma'} n(\Gamma,\gamma,\Gamma') ~\Gamma',
\end{equation}
which also yields the following formula for the coproduct,
\begin{equation}
\label{eq:cop-comb}
\Delta(\Gamma) = \sum_{\gamma,\Gamma'} \frac{n(\Gamma,\gamma,\Gamma')}{|\gamma|_\vee} ~  \gamma \otimes\Gamma'.
\end{equation}
\begin{rem}
From this last formula, one easily derives the Lie bracket on Feynman graphs as derived in \cite{CK99}. Indeed, one can define a pre-Lie product between 1PI graphs $\Gamma_1, \Gamma_2$ by duality 
$$
\langle \Gamma_1 \ast \Gamma_2 ,\Gamma \rangle := \langle \Gamma_1 \otimes \Gamma_2, \Delta(\Gamma) \rangle,
$$
with the pairing given by $\langle \Gamma_1, \Gamma_2 \rangle = 1$ if $\Gamma_1 \isom \Gamma_2$ and zero otherwise. This pre-Lie product defines a Lie bracket by $[\Gamma_1,\Gamma_2]= \Gamma_1 \ast \Gamma_2 - \Gamma_2 \ast \Gamma_1$ with $\ast$ given explicitly by
$$
\Gamma_1 \ast \Gamma_2 = \sum_\Gamma \frac{n(\Gamma, \Gamma_1, \Gamma_2)}{|\Gamma_1|_\vee}  ~ \Gamma. 
$$
\end{rem}

\begin{lma}
\label{lma:cop-gamma}
If $\Gamma$ and $\gamma$ are nonempty (connected) 1PI graphs, then $$n(\Gamma \circ_{(x,\phi)} \gamma,\gamma,\Gamma) = M(x,\phi)N(x,\phi).$$
\end{lma}
\begin{proof}
We have to count the number of subgraphs $\gamma' \isom \gamma$ of $\Gamma \circ_{(x,\phi)} \gamma$ such that there is an isomorphism $(\Gamma \circ_{(x,\phi)} \gamma) / \gamma' \isom  \Gamma$. 

This isomorphism can be trivial in the sense that there exists an element in $\Aut(\Gamma \circ_{(x,\phi)} \gamma)$ mapping $\gamma'$ to $\gamma$. Otherwise, the existence of such a isomorphism implies that there is an isomorphism $\Gamma \circ_{(x,\phi)} \gamma \isom \Gamma \circ_{(x',\phi')} \gamma$, with $(x',\phi')$ the image in $\Gamma$ of $\res(\gamma')$ in the quotient $(\Gamma \circ_{(x,\phi)} \gamma) / \gamma'$; such an isomorphism maps $\gamma$ in $\Gamma \circ_{(x,\phi)} \gamma$ to a certain subgraph $\gamma'$ of $\Gamma$. Moreover, $[x,\phi]$ and $[x',\phi']$ are in disjoint $\Aut(\Gamma)$-orbits, since if $(x',\phi')=g(x,\phi)$, the isomorphism would be the composition of an element in $\Aut(\Gamma \circ_{(x,\phi)} \gamma)$ and an element in $\Aut(\Gamma)$. 

We claim that all subgraphs $\gamma'$ obtained in this way (for disjoint orbits) are all different subgraphs of $\Gamma$, and cannot be the image of $\gamma$ under the action of an element in $\Aut(\Gamma \circ_{(x,\phi)} \gamma)$. This would then lead to $M(x,\phi)N(x,\phi)$ many subgraphs $\gamma'$ of $ \Gamma \circ_{(x,\phi)} \gamma$ satisfying $( \Gamma \circ_{(x,\phi)} \gamma)/\gamma' \isom \Gamma$.

Let $[x,\phi], [x',\phi'], [x'',\phi''] \in X(\Gamma,\gamma)$ be in disjoint orbits and suppose that there are isomorphisms 
\begin{align*}
g' :  \Gamma \circ_{(x',\phi')} \gamma &\to  \Gamma \circ_{(x,\phi)} \gamma,\\
g'' :  \Gamma \circ_{(x'',\phi'')} \gamma &\to  \Gamma \circ_{(x,\phi)} \gamma,
\end{align*}
mapping $\gamma$ to subgraphs $\gamma'$ and $\gamma''$ in $\Gamma$, respectively. If $\gamma'$ and $\gamma''$ coincide (up to an isomorphism $h$), then the composition $(g'')^{-1} \circ h \circ g'$ gives an isomorphism from $\Gamma \circ_{(x',\phi')} \gamma$ to $ \Gamma \circ_{(x'',\phi'')} \gamma$ mapping $\gamma$ to itself. It therefore induces an element in $\Aut(\Gamma)$ that sends $[x',\phi']$ to $[x'',\phi'']$, which cannot be true. We conclude that $\gamma'$ and $\gamma''$ are different subgraphs of $\Gamma$.

On the other hand, if there is an element $\phi$ in $\Aut(\Gamma \circ_{(x,\phi)} \gamma)$ that maps $\gamma$ to such a subgraph $\gamma' \in  \Gamma$, the composition $\phi^{-1}\circ g' $ would map $ \Gamma \circ_{(x',\phi')} \gamma$ to $\Gamma \circ_{(x,\phi)} \gamma$ isomorphically, sending $\gamma$ to itself. Again, such a map must be induced by an element in $\Aut(\Gamma)$ mapping $[x,\phi]$ to $[x',\phi']$, contradicting our assumptions. 
\end{proof}

\begin{lma}
\label{lma:defect}
Let $\gamma, \gamma'$ be as above. Then,
\begin{equation*}
\Delta_{\gamma \gamma'} = \frac{1}{n(\gamma, \gamma')}  ~ \Delta_\gamma \Delta_{\gamma'} - \rho_{\gamma\gamma'},
\end{equation*}
where $\rho_{\gamma \gamma'}$ is defined by 
\begin{equation*}
\rho_{\gamma\gamma'} = \sum_{[y,\psi] \in X(\gamma, \gamma')} \frac{\Sym(\gamma \circ_{(y,\psi)} \gamma')}{\Sym(\gamma\gamma')} \Delta_{\gamma \circ_{(y,\psi)} \gamma'}.
\end{equation*}
\end{lma}
\begin{proof}
Consider $\Delta_{\gamma\gamma'}(\Gamma)$ on a 1PI graph $\Gamma$; if $\gamma$ and $\gamma'$ appear as disjoint subgraphs of $\Gamma$, this expression is given by $\Delta_\gamma \Delta_{\gamma'}(\Gamma)$, up to a factor of $n(\gamma, \gamma')$ which corrects for the overcounting. Indeed, let $\gamma_1, \ldots, \gamma_m$ denote all subgraphs of $\Gamma$ that are isomorphic to $\gamma$. If $m \geq n$, then
\begin{align*}
\Delta_{\gamma^{n+1}} (\Gamma) &= \sum_{\begin{smallmatrix}\{i_1, \ldots, i_{n+1} \} \\ \subset \{1, \ldots, m\}\end{smallmatrix}} \frac{1}{(n+1)!}  ~  \Gamma/\gamma_{i_1} \cdots \gamma_{i_{n+1}};\\
\Delta_{\gamma^n} \Delta_\gamma (\Gamma) &= \sum_{i=1}^m \sum_{\begin{smallmatrix}\{i_1, \ldots, i_{n} \} \\ \subset  \{1, \ldots, \hat{i},\ldots, m\}\end{smallmatrix}} \frac{1}{n!} ~  \Gamma/\gamma_i \gamma_{i_1} \cdots \gamma_{i_{n}},
\end{align*}
leading precisely to the factor $n(\gamma^n, \gamma) = n+1$. On the other hand, if $m<n$, then both terms vanish.

In the case that $\Gamma$ contains a subgraph $\tilde\gamma$ such that $\tilde \gamma/\gamma' \isom \gamma$, we find a discrepancy between the two terms which is given by the following sum,
\begin{equation*}
\rho_{\gamma\gamma'} (\Gamma) =  \frac{1}{n(\gamma,\gamma')} \sum_{\tilde\gamma \subset \Gamma,\tilde \gamma /\gamma' \isom \gamma} n(\tilde\gamma, \gamma', \gamma) ~\Gamma/\tilde\gamma.
\end{equation*}
Here $n(\tilde\gamma, \gamma', \gamma)$ is by definition the number of disjoint subgraphs of $\tilde \gamma$ that are isomorphic to $\gamma'$ and such that $\tilde\gamma / \gamma' \isom \gamma$, which do indeed all contribute to $\Delta_\gamma \Delta_{\gamma'}(\Gamma)$. We replace the above sum by a sum over insertion places of $\gamma'$ in $\gamma$, while correcting for the equivalent insertions. The latter correcting factor is given as the number of elements $[y',\phi'] \in X(\gamma,\gamma')$ such that $\gamma \circ_{(y',\phi')} \isom \gamma \circ_{(y,\phi)}$. Such an isomorphism can be induced by an element $g \in \Aut(\gamma)$, with $[y',\phi'] = g[y,\phi]$ but leaving $\gamma'$ untouched, leading to a factor of $|\Aut(\gamma)[y,\phi]|$. The number of isomorphisms $\gamma \circ_{(y',\phi')} \isom \gamma \circ_{(y,\phi)}$ that are not induced by such an element, is given precisely by the factor $N(y,\phi)$. Thus, on inserting the expression for $n(\gamma \circ_{(y,\psi)} \gamma', \gamma',\gamma)$ derived in Lemma \ref{lma:cop-gamma}, we infer that,
\begin{align*}
\rho_{\gamma\gamma'} (\Gamma) &= \frac{1}{n(\gamma,\gamma')}  \sum_{[y,\psi] \in X(\gamma,\gamma')} \frac{M(y,\psi)N(y,\psi)}{N(y,\psi) |\Aut(\gamma)[y,\psi]|} ~  \Delta_{\gamma\circ_{(y,\psi)} \gamma'} (\Gamma)\\
&= \sum_{[y,\psi] \in X(\gamma,\gamma')}  \frac{\Sym(\gamma \circ_{(y,\psi)} \gamma')}{\Sym(\gamma\gamma')} ~ \Delta_{\gamma\circ_{(y,\psi)} \gamma'}(\Gamma),
\end{align*}
where we have applied Lemma \ref{lma:orbit} in going to the second line. 
We have also used Equation \eqref{eq:sym-union} to replace $ n(\gamma,\gamma') \Sym(\gamma) \Sym(\gamma')$ by $\Sym(\gamma\gamma')$. 
\end{proof}

\subsection{The coproduct on 1PI Green's functions}
\label{sect:coproduct}
Our main result of this section is the following.
\begin{prop}
\label{prop:linear}
The coproduct takes the following form on the 1PI Green's functions:
\begin{equation*}
\sum_{\Gamma^r_L}  \frac{1}{\Sym(\Gamma)} ~  \Delta(\Gamma) = \sum_{K=0}^L \sum_{\gamma_K,\Gamma_{L-K}^r} 
\frac{\ins{\Gamma}{\gamma}}{\Sym(\gamma)\Sym(\Gamma)}  ~ \gamma \otimes \Gamma,
\end{equation*}
where the sums are over all 1PI graphs $\Gamma$ with the indicated residue and loop number, and graphs $\gamma$ at the indicated loop order that are disjoint unions of 1PI graphs.
\end{prop}
\begin{proof}
Since $\Delta = \frac{1}{|\gamma|_\vee} \sum_\gamma \gamma \otimes \Delta_\gamma$, this would follow from the following equality, for $\gamma$ any disjoint union of 1PI graphs at loop order $K<L$ and $\gamma_0$ an auxiliary graph,
\begin{equation}
\label{eq:cop-gamma}
\sum_{\Gamma_L^r} \frac{\ins{\Gamma}{\gamma_0}}{|\gamma|_\vee\Sym(\Gamma)} ~  \Delta_\gamma (\Gamma) = \hspace{-2mm} \sum_{\Gamma_{L-K}^r} \hspace{-2mm} \frac{\ins{\Gamma}{\ins{\gamma}{\gamma_0}}}{\Sym(\gamma)\Sym(\Gamma)} ~ \Gamma.
\end{equation}
Indeed, putting $\gamma_0= \emptyset$ and summing over $\gamma$ then gives the desired result.
We show that Equation \eqref{eq:cop-gamma} holds by induction on the number of connected components of $\gamma$. 
\begin{lma}
\label{lma:ind1}
If $\gamma$ is a 1PI graph, then Equation \eqref{eq:cop-gamma} holds.
\end{lma}
\begin{proof}
If $\gamma=\emptyset$, there is nothing to prove, since $\Delta_\gamma(\Gamma)= \Gamma$, $\Sym(\emptyset)=1$ and $\ins{\Gamma}{\ins{\emptyset}{\gamma_0}} \equiv \ins{\Gamma}{\gamma_0}$. 
We claim that the following equality holds for $\gamma,\tilde\Gamma \neq \emptyset$,
\begin{equation*}
\sum_{\Gamma} \frac{\ins{\Gamma}{\gamma_0}}{|\gamma|_\vee\Sym(\Gamma)}  ~ n(\Gamma, \gamma, \tilde\Gamma) = \hspace{-5mm} \sum_{[x,\phi]\in X(\tilde\Gamma,\gamma)} \hspace{-1mm} \frac{\ins{\tilde\Gamma \circ_{(x,\phi)} \gamma}{\gamma_0} }{|\gamma|_\vee N(x,\phi) | \Aut(\tilde\Gamma)[x,\phi] |\Sym(\tilde\Gamma \circ_{(x,\phi)} \gamma)} ~ n(\tilde \Gamma \circ_{(x,\phi)} \gamma, \gamma, \tilde\Gamma).
\end{equation*}
Indeed, one can replace the sum on the left-hand-side over $\Gamma$ by a sum over insertion places of $\gamma$ in $\tilde\Gamma$ (so that $\Gamma \isom \tilde \Gamma \circ_{(x,\phi)} \gamma$ for some $[x,\phi] \in X(\Gamma,\gamma)$, and also $\res(\tilde\Gamma) = \res(\Gamma)$), provided one divides by a combinatorial factor counting the number of equivalent insertions. This factor is given as the number of elements $[x',\phi'] \in X(\Gamma,\gamma)$ such that $\tilde \Gamma \circ_{(x',\phi')} \gamma \isom \tilde\Gamma \circ_{(x,\phi)} \gamma$, in which case $\Sym(\tilde \Gamma \circ_{(x,\phi)} \gamma) = \Sym(\tilde \Gamma \circ_{(x',\phi')} \gamma)$ and also $\ins{\tilde\Gamma \circ_{(x,\phi)} \gamma}{\gamma_0} = \ins{\tilde\Gamma \circ_{(x',\phi')} \gamma}{\gamma_0}$.

Such an isomorphism $\tilde \Gamma \circ_{(x',\phi')} \gamma \isom \tilde\Gamma \circ_{(x,\phi)} \gamma$ can be induced by an element in $g \in \Aut(\tilde\Gamma)$ with $[x',\phi'] =g[x,\phi]$ but leaving $\gamma$ untouched. This leads to division by the length of the orbit $\Aut(\tilde\Gamma)[x,\phi]$. Otherwise, an isomorphism from $\tilde \Gamma \circ_{(x,\phi)} \gamma$ to $\tilde \Gamma \circ_{(x',\phi')} \gamma$ has to map $\gamma$ to an isomorphic subgraph $\gamma' \subset \tilde \Gamma$. In that case, it can not be induced by an element in $\Aut(\tilde\Gamma)$, leading precisely to the additional factor of $N(x,\phi)$.

Equation \eqref{eq:cop-gamma} now follows directly by inserting the expressions obtained in Lemma \ref{lma:orbit} and \ref{lma:cop-gamma} in the above equation and summing over all 1PI graphs $\tilde\Gamma$, as in Equation \eqref{eq:cop-gamma-comb}. We also noted on the way that by definition
$$
\frac{1}{|\gamma|_\vee}\sum_{[x,\phi]\in X(\tilde\Gamma,\gamma)} \ins{\tilde\Gamma \circ_{(x,\phi)} \gamma}{\gamma_0} = \ins{\tilde\Gamma}{\ins{\gamma}{\gamma_0}}.
$$
The case $\tilde\Gamma=\emptyset$ arises whenever $K=L$ and $\gamma \isom \Gamma$, 
in which case the combinatorial factors $\ins{\Gamma}{\gamma_0}$ and $\ins{\emptyset}{\ins{\gamma}{\gamma_0}}$ coincide.
\end{proof}

Assume now that Equation \eqref{eq:cop-gamma} holds for $\gamma$ a (non-empty) disjoint union of 1PI graphs of loop order $K$.
We will prove that it also holds for the disjoint union $\gamma \gamma' = \gamma \cup \gamma'$ of it with a non-empty 1PI graph $\gamma'$ of loop order $K'$. 
An application of Lemma \ref{lma:defect} yields,
\begin{align}
\label{eq:lhs}
\frac{\ins{\Gamma}{\gamma_0}}{|\gamma\gamma'|_\vee\Sym(\Gamma)} ~  \Delta_{\gamma\gamma'} (\Gamma) 
= 
\frac{\ins{\Gamma}{\gamma_0}}{n(\gamma,\gamma')|\gamma\gamma'|_\vee\Sym(\Gamma)} ~  \Delta_\gamma \Delta_{\gamma'}(\Gamma) 
-  \frac{\ins{\Gamma}{\gamma_0}}{|\gamma\gamma'|_\vee\Sym(\Gamma)} ~  \rho_{\gamma\gamma'}(\Gamma). 
\end{align}
Since $\gamma'$ is a 1PI graph, we can apply Lemma \ref{lma:ind1} to the first term, which gives for the sum over all graphs $\Gamma_L^r$,
\begin{align*}
\frac{1}{n(\gamma, \gamma')} \sum_{\Gamma_L^r} \frac{\ins{\Gamma}{\gamma_0}}{|\gamma\gamma'|_\vee\Sym(\Gamma)} ~  \Delta_\gamma \Delta_{\gamma'}(\Gamma) 
&= \frac{1}{n(\gamma, \gamma')} \sum_{\Gamma_{L-K'}^r} \frac{\ins{\Gamma}{\ins{\gamma'}{\gamma_0}}}{|\gamma|_\vee\Sym(\gamma') \Sym(\Gamma)}  ~ \Delta_\gamma(\Gamma)\\
&= \frac{1}{n(\gamma, \gamma')} \sum_{\Gamma_{L-K'}^r} \frac{\ins{\Gamma}{\gamma'\gamma_0} + (\ins{\Gamma}{\gamma'}) ( \ins{\gamma'}{\gamma_0})}{\Sym(\gamma') |\gamma|_\vee\Sym(\Gamma)}  ~ \Delta_\gamma(\Gamma),
\end{align*}
using also Equation \eqref{eq:ins3-ins}. The induction hypothesis -- that is, validity of Eq. \eqref{eq:cop-gamma} in the case of $\gamma$ -- now yields,
\begin{equation*}
\frac{1}{n(\gamma, \gamma')} \sum_{\Gamma_L^r} \frac{\ins{\Gamma}{\gamma_0}}{|\gamma\gamma'|_\vee\Sym(\Gamma)}  ~ \Delta_\gamma \Delta_{\gamma'}(\Gamma) 
=  \sum_{\Gamma_{L-K-K'}^r} \frac{\ins{\Gamma}{\ins{\gamma}{\gamma'\gamma_0}}+(\ins{\Gamma}{\ins{\gamma}{\gamma'}})(\ins{\gamma'}{\gamma_0})}{\Sym(\gamma\gamma') \Sym(\Gamma)}  ~ \Gamma,
\end{equation*}
combining once more the symmetry factors $\Sym(\gamma)$ and $\Sym(\gamma')$ with the help of $n(\gamma,\gamma')$. 
For the second term in Equation \eqref{eq:lhs}, we can use the induction hypothesis on $\Delta_{\gamma \circ_{(y,\psi)} \gamma'}$ to show that
\begin{align*}
\sum_{\Gamma_L^r} \frac{\ins{\Gamma}{\gamma_0}}{|\gamma\gamma'|_\vee\Sym(\Gamma)}  ~ \rho_{\gamma\gamma'}(\Gamma) 
= \sum_{\Gamma_{L-K-K'}^r} \sum_{[y,\psi] \in X(\gamma,\gamma')} \frac{\ins{\Gamma}{\ins{\gamma\circ_{(y,\psi)} \gamma'}}{\gamma_0}}{|\gamma'|_\vee\Sym(\gamma\gamma') \Sym(\Gamma)}  ~ \Gamma,
\end{align*}
since $|\gamma \circ_{(y,\psi)} \gamma'|_\vee = |\gamma|_\vee$. 
We conclude with the following equality, 
\begin{equation*}
\ins{\Gamma}{\ins{\gamma\gamma'}{\gamma_0}}=
\ins{\Gamma}{\ins{\gamma}{\gamma'\gamma_0}}+(\ins{\Gamma}{\ins{\gamma}{\gamma'}})(\ins{\gamma'}{\gamma_0}) - \frac{1}{|\gamma'|_\vee} \sum_{[y,\psi] \in X(\gamma, \gamma')}\ins{\Gamma}{\ins{\gamma\circ_{(y,\psi)} \gamma'}{\gamma_0}},
\end{equation*}
which follows easily from Definition \ref{defn:ins3}. Indeed, by definition
\begin{align*}
\ins{\Gamma}{\ins{\gamma}{\gamma'\gamma_0}}+(\ins{\Gamma}{\ins{\gamma}{\gamma'}})(\ins{\gamma'}{\gamma_0}) = \frac{1}{|\gamma\gamma'|_\vee} \sum_{[x,\phi] \in X(\Gamma, \gamma)} \sum_{[x',\phi'] \in X(\Gamma \circ_{(x,\phi)} \gamma, \gamma')} \hspace{-2mm} \ins{(\Gamma \circ_{(x,\phi)} \gamma) \circ_{(x',\phi')} \gamma'}{\gamma_0},
\end{align*}
which counts the number of places to insert $\gamma\gamma'$ and then $\gamma_0$ in $\Gamma$. Subtraction of the number of such places with $\gamma'$ sitting inside $\gamma$, leads precisely to the number of places to subsequently insert $\gamma \gamma'$ and $\gamma_0$ in $\Gamma$.
\end{proof}

\section{Ward identities in QED}
\label{sect:ward}
This section will be a warming-up of what is to come in the next section concerning non-abelian gauge theories. Quantum electrodynamics is an abelian gauge theory, and as a consequence, the Slavnov--Taylor identities (cf. Def. \ref{defn:st} below) become much more simple. More precisely, they become linear in the graphs and also known as Ward identities \cite{War50} (see Defn. \ref{defn:ward} below). 

We first make some observations about Feynman graphs in QED. We then proceed to prove compatibility of these Ward identities with the Hopf algebra of renormalization. 

\subsection{Feynman graphs in QED}
\label{subsect:graphs-qed}
In (massless) quantum electrodynamics, there is only the vertex of valence three, describing the interaction of the photon with a pair of electrons. There are two types of edges corresponding to the photon (wiggly edge) and the electron (straight edge). Summarizing, we have in the notation of the previous section: $R=R_V \cup R_E$ with 
\begin{align*}
R_V &=\{~ 
\parbox{20pt}{
    \begin{fmfgraph}(20,10)
      \fmfleft{l}
      \fmfright{r1,r2}
      \fmf{photon}{l,v}
      \fmf{plain}{r1,v}
      \fmf{plain}{v,r2}
    \end{fmfgraph}}~\};
\\
R_E &= \{~
\parbox{20pt}{
    \begin{fmfgraph}(20,10)
      \fmfleft{l}
      \fmfright{r}
      \fmf{plain}{l,r}
    \end{fmfgraph}}~,~
\parbox{20pt}{
    \begin{fmfgraph}(20,10)
      \fmfleft{l}
      \fmfright{r}
      \fmf{photon}{l,r}
    \end{fmfgraph}}~\}.
\end{align*}
In particular, this means that in the process of renormalization, only three types of graphs are of importance: the vertex graph, the electron self-energy graph and the vacuum polarization. Correspondingly, we define the following 1PI Green's functions,
\begin{align*}
G^\vertex &= 1+\sum_{\Gamma^\vertex} \frac{\Gamma}{\Sym(\Gamma)};\\
G^e &=1- \sum_{\Gamma^e} \frac{\Gamma}{\Sym(\Gamma)},
\end{align*}
with $e= \el,\ph$, and where the sum is over all 1PI Feynman graphs with the indicated residue. When this sum is restricted to 1PI graphs with loop number $L$, we denote this {\it Green's function at loop order $L$} by $G_L^r$. In particular, $G_0^r$ is understood as the 1PI graph with loop number zero, which is the empty graph; hence $G_0^r=1$. 
\begin{rem}
The (regularized) Feynman rules can be understood as an algebra map $\phi$ from $H$ to the field $K$ of Laurent series in the regularization parameter \cite{CK99}. Under this map, the above Green's functions are mapped to the corresponding Feynman amplitudes. In particular, we see that the (unrenormalized) effective action can be written as:
\begin{equation*}
S_\eff = \int - \frac{1}{4} \phi(G^\ph) F_{\mu\nu}^2 
 + \phi(G^\el) \bar\psi \gamma^\mu\partial_\mu \psi + e \phi(G^\vertex) \bar \psi A_\mu \psi. 
\end{equation*}
\end{rem}
A simplification in QED, is that the number $\ins{\Gamma}{\gamma}$ of insertion places only depends on the number of loops, the residue of $\Gamma$ and $\gamma$. Let $N_\el(r)$ denote the number of electron lines of $r$ and $N_\ph(r)$ the number of photon lines. Thus, if $\Gamma$ has residue $r$, then $N_\el(r)$ is the number of external electron lines of $\Gamma$ and $N_\ph(r)$ the number of its external photon lines. Also recall the notation $m_{\Gamma,r'}$ for the number of vertices/edges in $\Gamma$ of type $r'$, introduced just below Definition \ref{defn:ins}. 
The number of loops $L$ of the graph $\Gamma$ is then given by (see for instance \cite[Section 10.1]{PS95}), 
\begin{equation}
\label{eq:loops}
L= m_{\Gamma,\el} + m_{\Gamma,\ph} - m_{\Gamma, \vertex} + 1,
\end{equation}
where the number of vertices $m_{\Gamma,\vertex}$ in $\Gamma$ with residue $r$ can be expressed as, 
\begin{align*}
m_{\Gamma,\vertex}&=2 m_{\Gamma, \ph} + N_\ph(r)=m_{\Gamma, \el} + \half N_\el(r).
\end{align*}
Combining these equalities, we arrive at the following relations,
\begin{align*}
m_{\Gamma, \vertex} &= 2 L + N_\el(r)+N_\ph(r) - 2;\\
m_{\Gamma, \ph} &= L + \half N_\el(r)  - 1;\\
m_{\Gamma, \el} &= 2 L +  \half N_\el(r)+N_\ph(r) - 2 ,
\end{align*}
from which we infer that the number $\ins{\Gamma}{\gamma}$ can be expressed in terms of $L$, $r$ and $\gamma$. We denote this number by $\ins{(L,r)}{\gamma}$. 

Explicitly, we have for example,
\begin{equation}
\label{rel:ins-qed}
\begin{aligned}
\ins{(L,\vertex)} {\gamma}
&= n_{\gamma,\vertex}! { 2L+1 \choose n_{\gamma,\vertex}} 
n_{\gamma,\el}! { 2L+n_{\gamma,\el} -1 \choose n_{\gamma,\el} }
n_{\gamma,\ph}! { L+n_{\gamma,\ph} -1 \choose n_{\gamma,\ph} },\\
\ins{(L,\el)}{\gamma}
&= n_{\gamma,\vertex}! { 2L \choose n_{\gamma,\vertex}}
n_{\gamma,\el}! { 2L+n_{\gamma,\el} -2 \choose n_{\gamma,\el} }
n_{\gamma,\ph}! { L+n_{\gamma,\ph} -1 \choose n_{\gamma,\ph} }.
\end{aligned}
\end{equation}

\subsection{The Ward identities}
In QED, there are relations between the (amplitudes of the) Green's functions $\phi(G^\vertex)$ and $\phi(G^\el)$; these relations play an important role in the process of renormalization. We will show here that they can be implemented on the Hopf algebra, in a way that is compatible with the coproduct. 
\begin{defn}
\label{defn:ward}
The {\rm Ward elements} $W_L$ at loop order $L>0$ are defined by
\begin{equation*}
W_L := G^\vertex_L - G^\el_L \equiv
\sum_{\Gamma_L^\vertex}
\frac{\Gamma}{\Sym(\Gamma)} + \sum_{\Gamma_L^\el} \frac{\Gamma}{\Sym(\Gamma)},
\end{equation*}
where the sum is over 1PI Feynman graphs with loop number $L$ and with the indicated residue. Moreover, we set $W_{L=0}:=G_{L=0}^\vertex - G_{L=0}^\el = 0$.
\end{defn}

Before stating our main result on the compatibility of the Ward elements with the coproduct, we introduce the following combinatorial factor,
\begin{equation}
\label{eq:c}
c(L,\gamma) := n_{\gamma,\vertex}! { 2L+1 \choose n_{\gamma,\vertex}} 
n_{\gamma,\el}! { 2L+n_{\gamma,\el} -1 \choose n_{\gamma,\el} }
n_{\gamma,\ph}! { L+n_{\gamma,\ph} -1 \choose n_{\gamma,\ph} },
\end{equation}
Note the mixture between the two factors $\ins{(L,\vertex)} {\gamma}$ and $\ins{(L,\el)} {\gamma}$ of Equation \eqref{rel:ins-qed} above. 
If the vertex graphs in $\gamma$ are labeled as $\gamma_1, \ldots, \gamma_{n_{\gamma,\vertex}}$, one could also define $c(L,\gamma)$ recursively by,
\begin{align}
\label{eq:c-rec}
\ins{(L,\vertex)}{\gamma\gamma_{n+1}} -  \sum_{l=1}^{n+1} c(L,\gamma \gamma_{n+1} - \gamma_l) = c(L,\gamma\gamma_{n+1}),
\end{align}
for a vertex graph $\gamma_{n+1}$ while setting $c(L,\gamma)=\ins{(L,\vertex)}{\gamma}$ if $n_{\gamma,\vertex}=0$.

\begin{thm}
\label{thm:ward}
For any $L \geq 0$, we have 
\begin{multline*}
\Delta(W_L) =  \hspace{-2mm}  \sum_{K+K'=0}^N  \hspace{-2mm} W_K \sum_{\gamma_{K'}} \frac{c(L-K-K',\gamma)}{\Sym(\gamma)} ~ \gamma \otimes G^\vertex_{L-K-K'} 
+ \sum_{K=0}^L \sum_{\gamma_{K}} \frac{\ins{(L-K,\el)}{\gamma}}{\Sym(\gamma)}~ \gamma \otimes W_{L-K}.
\end{multline*}
Consequently, the ideal $I$ generated by the Ward elements $W_L$ for every $L$ is a Hopf ideal in $H$,
\begin{gather*}
\Delta(I) \subseteq I \otimes H + H \otimes I ,\qquad \epsilon(I)=0,\qquad S(I) \subseteq I.
\end{gather*}
\end{thm}
\begin{proof}
Essential for the proof will be the relation between the two factors $\ins{\Gamma_L^\vertex}{\gamma}$ and $\ins{\Gamma_L^\el}{\gamma}$ displayed in Equation \eqref{rel:ins-qed}.
Using Pascal's rule, Eq. \eqref{pascal}, 
we derive the following relation between the two numbers:
\begin{multline*}
\label{eq:vertex-el}
\ins{\Gamma_L^\vertex}{\gamma} = \ins{\Gamma_L^\el}{\gamma} 
+ n_{\gamma,\vertex}! { 2L \choose n_{\gamma,\vertex}-1} 
n_{\gamma,\el}! { 2L+n_{\gamma,\el} -1 \choose n_{\gamma,\el} }
n_{\gamma,\ph}! { L+n_{\gamma,\ph} -1 \choose n_{\gamma,\ph} }
\\
+ n_{\gamma,\vertex}! { 2L \choose n_{\gamma,\vertex}}  
n_{\gamma,\el}! { 2L+n_{\gamma,\el} -2 \choose n_{\gamma,\el}-1 }
n_{\gamma,\ph}! { L+n_{\gamma,\ph} -1 \choose n_{\gamma,\ph} }.
\end{multline*}
Before inserting this in the expression for $\Delta(G^\vertex_L - G^\el_L)$ derived in Proposition \ref{prop:linear}, we observe that the second term is just $c(L,\tilde\gamma)$ with $\tilde \gamma$ the graph $\gamma$ with one vertex graph $\gamma_v$ subtracted, {\it times a factor of} $n_{\gamma,\vertex}=n_{\tilde\gamma,\vertex}+1$. Similarly, the third term is $c(L,\tilde\gamma)$ where now $\tilde\gamma$ is $\gamma$ with one electron self-energy graph $\gamma_e$ subtracted, {\it times a factor of} $n_{\gamma,\el}=n_{\tilde\gamma,\el}+1$. Note that in the respective cases $n_{\gamma,\vertex}=0$ and $n_{\gamma,\el}=0$, the above two terms vanish. We then find that,
\begin{align*}
\Delta( G^\vertex_L - G^\el_L)
&=  \sum_{K=0}^L \sum_{\gamma_{K}} \frac{\ins{(L-K,\el)}{\gamma}}{\Sym(\gamma)}~ \gamma \otimes \left( \sum_{\Gamma_{L-K}^\vertex} \frac{\Gamma}{\Sym(\Gamma)} + \sum_{\Gamma_{L-K}^\el} \frac{\Gamma}{\Sym(\Gamma)} \right)\\
&+ \sum_{K+K'}^L \sum_{\gamma_{e,K},\tilde\gamma_{K'}} \frac{ c(L-K-K',\tilde\gamma) }{\Sym(\gamma_e)\Sym(\gamma)} \gamma_e \tilde\gamma \otimes \sum_{\Gamma_{L-K-K'}^\vertex} \frac{\Gamma}{\Sym(\Gamma)}\\
&+\sum_{K+K'}^L \sum_{\gamma_{v,K},\tilde\gamma_{K'}} \frac{ c(L-K-K',\tilde\gamma) }{\Sym(\gamma_v)\Sym(\gamma)} \gamma_v \tilde\gamma \otimes \sum_{\Gamma_{L-K-K'}^\vertex} \frac{\Gamma}{\Sym(\Gamma)}.
\end{align*}
The above factors $n_{\gamma,\vertex}$ and $n_{\gamma,\el}$ are precisely canceled when replacing the sum over $\gamma$ by a sum over $\gamma_e$ and $\tilde\gamma$. Indeed, in doing so, a factor of $n(\tilde\gamma,\gamma_e)/(n_{\tilde\gamma,\el}+1)$ is needed to correct for overcounting,
\begin{equation*}
\sum_{\begin{smallmatrix}\gamma_e, \tilde\gamma \\ \gamma_e \tilde\gamma \isom \gamma \end{smallmatrix}}
\frac{n(\tilde\gamma,\gamma_e)}{n_{\tilde\gamma,\el}+1} ~ \frac{1}{\Sym(\gamma_e\tilde\gamma)} ~ F(\gamma_e\tilde\gamma)= \sum_{\gamma_e} \frac{n(\gamma,\gamma_e)-1}{n_{\gamma,\el}}~  \frac{1}{\Sym(\gamma)} ~ F(\gamma) =  \frac{1}{\Sym(\gamma)}~ F(\gamma),
\end{equation*}
by the very definition of $n(\gamma,\gamma_e)$, and a similar result holds for $\gamma_v$. We thus have respective factors of $n_{\gamma,\vertex}$ and $n_{\gamma,\el}$ in the denominator, and combining $n(\tilde\gamma,\gamma_e)$ with $\Sym(\gamma_e\tilde \gamma)$ (and similarly for the analogous expression for $\gamma_v$) using Equation \eqref{eq:sym-union} yields,
\begin{equation}
\label{eq:split-gamma}
\begin{aligned}
n_{\gamma,\el} \sum_\gamma \frac{1}{\Sym(\gamma)}~ F(\gamma) 
&=  \sum_{\gamma_e^\el,\tilde\gamma} \frac{1}{\Sym(\gamma_e)\Sym(\tilde\gamma)} ~F(\gamma_e\tilde\gamma); \\
n_{\gamma,\vertex} \sum_\gamma \frac{1}{\Sym(\gamma)}~ F(\gamma) 
&=  \sum_{\gamma_v^\vertex,\tilde\gamma} \frac{1}{\Sym(\gamma_v)\Sym(\tilde\gamma)} ~F(\gamma_v\tilde\gamma) .
\end{aligned}
\end{equation}
\end{proof}

\section{Non-abelian gauge theories}
\label{sect:st}
In this section, we come to the main purpose of this article, and show compatibility of the Slavnov--Taylor identities with the Hopf algebraic structure of renormalization of non-abelian gauge fields. Before that, we carefully describe the setting of non-abelian gauge theories. In particular, we start by describing the graphs that are allowed in such theories and list some combinatorial properties of the number $\ins{\Gamma}{\gamma}$ of insertion places defined in Section \ref{sect:ins}. These properties will be essential in the proof of compatibility of the Slavnov--Taylor identities with the coproduct.

\subsection{Feynman graphs in non-abelian gauge theories}
In order to make the following as concrete as possible, we work in the setting of the non-abelian gauge theory {\it quantum chromodynamics} (QCD). It describes the interaction between quarks (the fermions) via gluons (the gauge bosons).

In contrast with quantum electrodynamics described previously, there are now three vertices of valence three, describing the interaction of the fermion and ghost with the gluon, as well as the cubic gluon self-interaction. In addition, there is the quartic gluon self-interaction. This means that the Feynman graphs are built from the following two sets of vertices and edges:
\begin{align*}
R_V &= \{ 
\raisebox{-7.5pt}{
\parbox{20pt}{
    \begin{fmfchar}(20,15)
      \fmfleft{l}
      \fmfright{r1,r2}
      \fmf{gluon}{l,v}
      \fmf{plain}{r1,v}
      \fmf{plain}{v,r2}
    \end{fmfchar}
  }}
,
\raisebox{-7.5pt}{
\parbox{20pt}{
  \begin{fmfgraph}(20,15)
      \fmfleft{l}
      \fmfright{r1,r2}
      \fmf{gluon}{l,v}
      \fmf{dots}{r1,v}
      \fmf{dots}{v,r2}
  \end{fmfgraph}
}}
,
\raisebox{-7.5pt}{
\parbox{20pt}{
  \begin{fmfgraph}(20,15)
    \fmfleft{l}
      \fmfright{r1,r2}
      \fmf{gluon}{l,v}
      \fmf{gluon}{r1,v}
      \fmf{gluon}{v,r2}
  \end{fmfgraph}
}}
,
\raisebox{-7.5pt}{
\parbox{20pt}{
  \begin{fmfgraph}(20,15)
    \fmfleft{l1,l2}
      \fmfright{r1,r2}
      \fmf{gluon}{l1,v}
      \fmf{gluon}{l2,v}
      \fmf{gluon}{r1,v}
      \fmf{gluon}{v,r2}
  \end{fmfgraph}
}}
\};\\
R_E &= \{ 
\raisebox{-7.5pt}{
\parbox{20pt}{
  \begin{fmfgraph}(20,10)
      \fmfleft{l}
      \fmflabel{}{l}
      \fmfright{r}
      \fmf{plain}{l,r}
  \end{fmfgraph}
}}
,
\raisebox{-7.5pt}{
\parbox{20pt}{
  \begin{fmfgraph}(20,10)
      \fmfleft{l}
      \fmflabel{}{l}
      \fmfright{r}
      \fmf{dots}{l,r}
  \end{fmfgraph}
}}
,
\raisebox{-7.5pt}{
\parbox{20pt}{
  \begin{fmfgraph}(20,10)
      \fmfleft{l}
      \fmflabel{}{l}
      \fmfright{r}
      \fmf{gluon}{l,r}
  \end{fmfgraph}
}}
\},
\end{align*}
where the plain, dotted and curly lines represent the quark, ghost and gluon, respectively. Corresponding to the elements in $R$, we define 7 (1PI) Green's functions,
\begin{align*}
G^v &= 1 + \sum_{\Gamma^v} \frac{\Gamma}{\Sym(\Gamma)}  \qquad (v \in R_V);\\
G^e &= 1 - \sum_{\Gamma^e} \frac{\Gamma}{\Sym(\Gamma)}  \qquad (e \in R_E),
\end{align*}
with the sum over all 1PI Feynman graphs with the indicated residue, {i.e.} $\res(\Gamma^r)=r$.
\begin{rem}
\label{rem:eff-action}
As for quantum electrodynamics, the (regularized) Feynman rules for QCD (as listed for instance on page 34 of \cite{Col84}) can be understood as an algebra map $\phi$ from $H$ to the field $K$ of Laurent series in the regularization parameter \cite{CK99}. The (unrenormalized) effective action can be written as,
\begin{align*}
S_\eff &=\int -\frac{1}{4} \phi(G^\glu) \left( \partial_\mu A^a_\nu - \partial_\nu A^a_\mu \right)^2 
+ \phi(G^\qua) \bar\psi \gamma^\mu\partial_\mu \psi 
+ \phi(G^\gho) \partial_\mu \bar c \partial_\mu c 
+ g \phi(G^\quaglu) \bar \psi A_\mu \psi \\
&\quad+  g \phi(G^\quaglu) \bar \psi A_\mu \psi
 + g \phi(G^\gluc) \left( \partial_\mu A^a_\nu - \partial_\nu A^a_\mu \right) f_{abc} A^{b\mu} A^{c\nu}
+ g^2 \phi(G^\gluq) \left( f_{abc} A^b_\mu A^c_\nu \right)^2.
\end{align*}
\end{rem}

Contrary to the case of QED (cf. Sect. \ref{subsect:graphs-qed}), there are no relations in QCD expressing the numbers $m_{\Gamma,r}$ of vertices/edges of $\Gamma$ of the type $r$ in terms of the loop number and the residue of a graph $\Gamma$. 
However, we do have relations between them. 
Denote by $N_e(r)$ the number of lines attached to $r$ that are of the type $e \in R_E$; by convention, $N_e(e')=2 \delta_{e,e'}$ for an edge $e'$. Set $N(r) = \sum_e N_e(r)$, it is the total number of lines attached to $r$.
\begin{lma}
\label{lma:m}
Let $\Gamma$ be a (QCD) 1PI Feynman graph at loop order $L$ and with residue $r$. Then, the following relations hold 
\begin{equation*}
\begin{aligned}
\sum_{e \in R_E} m_{\Gamma,e} - \sum_{v \in R_V} m_{\Gamma,v} + 1 &=L; \quad &{\rm (a)}
 \qquad & \qquad m_{\Gamma,\quaglu}=m_{\Gamma,\qua} + \half N_\qua(r); \quad &{\rm (b)}\\
 \sum_{\begin{smallmatrix}v \in R_V \\ \val ~v =3 \end{smallmatrix}} m_{\Gamma,v} + 
2 m_{\Gamma,\gluq} - N(r) + 2 &= 2L; \quad &{\rm (c)}   
\qquad &  \qquad m_{\Gamma,\ghoglu}=m_{\Gamma,\gho} + \half N_\gho(r). \quad &{\rm (d)}
\end{aligned}
\end{equation*}
\end{lma}
\begin{proof}
Equation (a) is the usual expression of the Betti number of the graph $\Gamma$; it can be proved by induction on the number of edges of the graph $\Gamma$. Equations (b) and (d) follow from the statement that each of these two vertices involve exactly two quark or two ghost lines, respectively. We prove that Equation (c) holds by induction on the number of edges. First, denote by $V_3$ and $V_4$ the number of all trivalent and quadrivalent vertices of $\Gamma$, respectively. Since the type of vertices and edges is irrelevant for Eq. (c), we have to show validity of, 
\begin{equation}
\label{eq:l-v}
V_3 + 2 V_4 - N + 2 = 2L,
\end{equation} 
for a graph $\Gamma$ consisting of $V_3$ trivalent and $V_4$ quadrivalent vertices of a single fixed type. Induction on the number of edges starts with the graph consisting of only one edge, for which $N=2$, by definition. Then, if we add an edge to a graph $\Gamma$ for which Eq. \eqref{eq:l-v} holds, we have the following possibilities to connect the new edge:
\begin{center}
\begin{tabular}{l|rrrr}
&$\delta V_3$ & $\delta V_4$ & $\delta N$ & $\delta L$\\
\hline
edge - edge & $+2$ & & & $+1$\\
edge - edge & & $+1$ & & $+1$\\
edge - 3 vertex &  & $+1$ & & $+1$\\
3 vertex - 3 vertex & $-2$ & $+2$ & & $+1$\\
edge - ext & $+1$ & & $+1$ & \\
3 vertex - ext & $-1$ & $+1$ & $+1$ & 
\end{tabular}
\end{center}
One readily checks that the corresponding changes in $V_3, V_4, N$ and $L$ leave Eq. \eqref{eq:l-v} invariant. 
\end{proof}
Motivated by these properties, we introduce the following notion of admissibility of a vector $\vec{m}=(m_r)_{r \in R}$.
\begin{defn}
A vector $\vec{m}=(m_r)_{r \in R}$ is called {\rm $(L,r)$-admissible} if it satisfies Equations (a)-(d) in Lemma \ref{lma:m}. 
\end{defn}

\begin{lma}
\label{lma:steps}
Two $(L,r)$-admissible vectors differ by a finite number of combinations of the following three steps
\begin{center}
\begin{tabular}{c|lllllll}
& $\delta m_\qua$ & $\delta m_\gho$ & $\delta m_\glu$ & $\delta m_\quaglu$ & $\delta m_\ghoglu$ & $\delta m_\gluc$ & $\delta m_\gluq$ \\
\hline
{\rm I} & $+1$ &&&$+1$&&$+1$ &$-1$\\
{\rm II} & & $+1$ &&& $+1$& $+1$& $-1$\\
{\rm III} & && $+1$ &&& $+2$ & $-1$
\end{tabular}
\end{center}
acting by $m_r \to m_r + \delta m_r$, while retaining admissibility at each step.
\end{lma}
\begin{proof}
Clearly, two admissible vectors differ by integers in each of the entries $m_e$ for $e \in R_V$. Since Equations (a)-(d) in Lemma \ref{lma:m} impose linear constraints on the entries of $m$, it is enough to consider the cases $m_e \to m_e +1$ for each of the three edges individually (thus leaving the other $m_e$ invariant). Then, Equations (a)-(d) become four independent constraints on the four remaining entries $m_v$, $v \in R_V$, the solution of which is displayed in the above table.
\end{proof}

\subsection{Slavnov--Taylor identities}
Contrary to the linear Ward identities that we have encountered in QED, there are now quadratic relations between Green's functions that reflect the non-abelian nature of the gauge symmetry of the corresponding classical field theory. 
For a derivation of the Slavnov--Taylor identities, we refer the reader to the standard text books on quantum field theory, such as \cite[Sect. 17.1]{Wei96} or \cite[Sect. 21.4-5]{Zin89}. Pictorially, we have the following three identities:
\begin{align*}
\tag{I}
\parbox{30pt}{ 
    \begin{fmfgraph*}(30,30)
      \fmfleft{l}
      \fmfright{r1,r2}
      \fmf{gluon}{l,v}
      \fmf{gluon}{r1,v}
      \fmf{gluon}{v,r2}
      \fmfblob{.5w}{v}
    \end{fmfgraph*}
}
\parbox{30pt}{ 
    \begin{fmfgraph*}(30,30)
    \fmfleft{l}
      \fmfright{r1,r2}
      \fmf{gluon}{l,v}
      \fmf{plain}{r1,v}
      \fmf{plain}{v,r2}
      \fmfblob{.5w}{v}
    \end{fmfgraph*}
}
-
\parbox{30pt}{ 
  \begin{fmfgraph*}(30,30)
    \fmfleft{l1,l2}
      \fmfright{r1,r2}
      \fmf{gluon}{l1,v}
      \fmf{gluon}{l2,v}
      \fmf{gluon}{r1,v}
      \fmf{gluon}{v,r2}
      \fmfblob{.5w}{v}
  \end{fmfgraph*}
}
\parbox{30pt}{ 
  \begin{fmfgraph*}(30,30)
      \fmfleft{l}
      \fmflabel{}{l}
      \fmfright{r}
      \fmf{plain}{l,v,r}
      \fmfblob{.5w}{v}
  \end{fmfgraph*}
}
&=0;
\\
\tag{II}
\parbox{30pt}{ 
    \begin{fmfgraph*}(30,30)
      \fmfleft{l}
      \fmfright{r1,r2}
      \fmf{gluon}{l,v}
      \fmf{gluon}{r1,v}
      \fmf{gluon}{v,r2}
      \fmfblob{.5w}{v}
    \end{fmfgraph*}
}
\parbox{30pt}{ 
    \begin{fmfgraph*}(30,30)
    \fmfleft{l}
      \fmfright{r1,r2}
      \fmf{gluon}{l,v}
      \fmf{dots}{r1,v}
      \fmf{dots}{v,r2}
      \fmfblob{.5w}{v}
    \end{fmfgraph*}
}
-
\parbox{30pt}{ 
  \begin{fmfgraph*}(30,30)
    \fmfleft{l1,l2}
      \fmfright{r1,r2}
      \fmf{gluon}{l1,v}
      \fmf{gluon}{l2,v}
      \fmf{gluon}{r1,v}
      \fmf{gluon}{v,r2}
      \fmfblob{.5w}{v}
  \end{fmfgraph*}
}
\parbox{30pt}{ 
  \begin{fmfgraph*}(30,30)
      \fmfleft{l}
      \fmflabel{}{l}
      \fmfright{r}
      \fmf{dots}{l,v,r}
      \fmfblob{.5w}{v}
  \end{fmfgraph*}
}
&=0;
\\
\tag{III}
\parbox{30pt}{ 
    \begin{fmfgraph*}(30,30)
      \fmfleft{l}
      \fmfright{r1,r2}
      \fmf{gluon}{l,v}
      \fmf{gluon}{r1,v}
      \fmf{gluon}{v,r2}
      \fmfblob{.5w}{v}
    \end{fmfgraph*}
}
\parbox{30pt}{ 
    \begin{fmfgraph*}(30,30)
    \fmfleft{l}
      \fmfright{r1,r2}
      \fmf{gluon}{l,v}
      \fmf{gluon}{r1,v}
      \fmf{gluon}{v,r2}
      \fmfblob{.5w}{v}
    \end{fmfgraph*}
}
-
\parbox{30pt}{ 
  \begin{fmfgraph*}(30,30)
    \fmfleft{l1,l2}
      \fmfright{r1,r2}
      \fmf{gluon}{l1,v}
      \fmf{gluon}{l2,v}
      \fmf{gluon}{r1,v}
      \fmf{gluon}{v,r2}
      \fmfblob{.5w}{v}
  \end{fmfgraph*}
}
\parbox{30pt}{ 
  \begin{fmfgraph*}(30,30)
      \fmfleft{l}
      \fmflabel{}{l}
      \fmfright{r}
      \fmf{gluon}{l,v,r}
      \fmfblob{.5w}{v}
  \end{fmfgraph*}
}
&=0,
\end{align*}
where the blob stands for the 1PI Green's function corresponding to the indicated external structure.
In the Hopf algebraic setting of renormalization, this motivates the following definition. 
\begin{defn}
\label{defn:st}
The {\rm Slavnov--Taylor elements} are defined by
\begin{align*}
&\ST^\I = G^\gluc G^\quaglu - G^\gluq G^\qua; \\
&\ST^\II = G^\gluc G^\ghoglu - G^\gluq G^\gho;\\
&\ST^\III= G^\gluc G^\gluc - G^\gluq G^\glu. 
\end{align*}
\end{defn}
Note that these elements involve both a linear and quadratic part, for instance,\begin{align*}
\ST^\I &= \sum_{\Gamma^\gluc} \frac{\Gamma}{\Sym(\Gamma)} + \sum_{\Gamma^\quaglu} \frac{\Gamma}{\Sym(\Gamma)} + \sum_{\Gamma_1^\gluc,\Gamma_2^\quaglu} \frac{\Gamma_1 \Gamma_2}{\Sym(\Gamma_1)\Sym(\Gamma_2)} \\
&\qquad- \sum_{\Gamma^\gluq} \frac{\Gamma}{\Sym(\Gamma)} + \sum_{\Gamma^\qua} \frac{\Gamma}{\Sym(\Gamma)} + \sum_{\Gamma_1^\gluq,\Gamma_2^\qua} \frac{\Gamma_1 \Gamma_2}{\Sym(\Gamma_1)\Sym(\Gamma_2) }~.
\end{align*}
The choice for using the same labelling $\I,~\II,~\III$ for the Slavnov--Taylor identities as for the admissible steps in Lemma \ref{lma:steps} is not coincidental, but motivated by the next Lemma. First, define for vectors $\vec{m}=(m)_{r\in R}$ and $\vec{n}=(n)_{r\in R}$ the following constant,
\begin{equation}
c \begin{pmatrix} \vec{m}\\ \vec{n} \end{pmatrix}=\prod_{v \in R_V} n_v! {m_v  \choose n_v} \prod_{e \in R_E} n_e! {m_e+n_e-1  \choose n_e}.
\end{equation}
A glance back at Section \ref{sect:ins}, makes one realize that whenever $\vec{m}$ and $\vec{n}$ arise from two graphs $\Gamma$ and $\gamma$ ({i.e.} $\vec{m}=\vec{m}_\Gamma, \vec{n}=\vec{n}_\gamma$), this constant becomes the 
number $\ins{\Gamma}{\gamma}$ of insertion places of $\gamma$ in $\Gamma$. 
We also introduce the following standard basis $\{ \vec{f}_r\}_{r \in R}$ of vectors corresponding to the elements $r \in R$ by setting $(f_r)_{r'} = \delta_{rr'}$.
\begin{lma}
\label{lma:steps-st}
The following equation holds for $A=\I,~\II,~\III$,
\begin{align*}
\sum_\gamma \left[c \begin{pmatrix} \vec{m}+\delta \vec{m}^A \\ \vec{n} \end{pmatrix} -  c \begin{pmatrix} \vec{m}\\ \vec{n} \end{pmatrix}  \right] &= \ST^A ~ \sum_\gamma c \begin{pmatrix} \vec{m} - \vec{f}_\gluq \\ \vec{n} \end{pmatrix} \frac{\gamma}{\Sym(\gamma)}~.
\end{align*}
\end{lma}
\begin{proof}
First observe that the cases $\I$ and $\II$ can be treated simultaneously by exchanging external electron lines with external ghost lines. After applying Pascal's rule \eqref{pascal} eight times, we obtain
\begin{multline*}
c \begin{pmatrix} \vec{m}+\delta \vec{m}^\I \\ \vec{n} \end{pmatrix}  - c \begin{pmatrix} \vec{m}\\ \vec{n} \end{pmatrix} = 
n_\qua ~c \begin{pmatrix} \vec{m} - \vec{f}_\gluq \\ \vec{n} - \vec{f}_\qua \end{pmatrix}
- n_\gluq ~c \begin{pmatrix} \vec{m} - \vec{f}_\gluq \\ \vec{n} - \vec{f}_\gluq \end{pmatrix}
+n_\gluq n_\qua ~c \begin{pmatrix} \vec{m} - \vec{f}_\gluq \\ \vec{n} - \vec{f}_\gluq - \vec{f}_\qua \end{pmatrix}\\
+n_\gluc ~c \begin{pmatrix} \vec{m} - \vec{f}_\gluq \\ \vec{n} - \vec{f}_\gluc \end{pmatrix}
+n_\quaglu~ c \begin{pmatrix} \vec{m} - \vec{f}_\gluq \\ \vec{n} - \vec{f}_\quaglu \end{pmatrix}
+n_\quaglu n_\gluc~ c \begin{pmatrix} \vec{m} - \vec{f}_\gluq \\ \vec{n} - \vec{f}_\quaglu - \vec{f}_\gluc  \end{pmatrix}.
\end{multline*}
Now, the analogue of Equation \eqref{eq:split-gamma} in the setting of QCD yields, when applied to these six terms, precisely the above formula in the case $A=\I$ (and similarly for $A=\II$). 

In like manner, one shows that 
\begin{multline*}
c \begin{pmatrix} \vec{m}+\delta \vec{m}^\III \\ \vec{n} \end{pmatrix}  - c \begin{pmatrix} \vec{m}\\ \vec{n} \end{pmatrix} = 
n_\glu ~c \begin{pmatrix} \vec{m} - \vec{f}_\gluq \\ \vec{n} - \vec{f}_\glu \end{pmatrix}
- n_\gluq ~c \begin{pmatrix} \vec{m} - \vec{f}_\gluq \\ \vec{n} - \vec{f}_\gluq \end{pmatrix}
+n_\gluq n_\glu ~c \begin{pmatrix} \vec{m} - \vec{f}_\gluq \\ \vec{n} - \vec{f}_\gluq - \vec{f}_\glu \end{pmatrix}\\
+2 n_\gluc ~c \begin{pmatrix} \vec{m} - \vec{f}_\gluq \\ \vec{n} - \vec{f}_\gluc \end{pmatrix}
+n_\gluc (n_\gluc-1) ~ c \begin{pmatrix} \vec{m} - \vec{f}_\gluq \\ \vec{n} - 2\vec{f}_\gluc  \end{pmatrix},
\end{multline*}
from which, with the proper analogue of Eq. \eqref{eq:split-gamma}, one concludes the above formula for $A=\III$.
\end{proof}
The above Lemma will be essential in the proof of our main result, stating compatibility between the Slavnov--Taylor elements and the Hopf algebra structure of renormalization.
\begin{thm}
\label{thm:st}
The ideal $I$ generated by the Slavnov--Taylor elements is a Hopf ideal, {i.e.} 
\begin{gather*}
\Delta(I) \subseteq I \otimes H + H\otimes I, \qquad \epsilon(I)=0, \qquad S(I) \subseteq I.
\end{gather*}
\end{thm}
\begin{proof}
By the very definition of the ideal $I$ the second condition is trivially satisfied whereas the third follows from the first, by the recursive formula for the antipode in Equation \eqref{antipode}. It is thus enough to show that $\Delta(\ST^A) \subseteq I \otimes H + H \otimes I$ for $A=\I,~\II,~\III$. Clearly, the primitive part of the coproduct $\Delta(\ST^A)$ is of the desired form, by the very definition of being the primitive part.
We will apply Proposition \ref{prop:linear} to determine the non-primitive part. Starting with the second term in $\ST^\I$, we find that,
\begin{align}
\label{cop:gluq-qua}
\Delta'\left(G^\gluq G^\qua \right) &= 
 \hspace{-2mm} \sum_{\gamma,\Gamma_1^\gluq \neq \emptyset}  \hspace{-2mm} \frac{\ins{\Gamma_1}{\gamma}}{\Sym(\gamma) \Sym(\Gamma_1)} ~\gamma \otimes \Gamma_1 
-  \hspace{-3mm} \sum_{\gamma,\Gamma_1^\gluq,\Gamma_2^\qua\neq \emptyset} \hspace{-2mm} \frac{\ins{\Gamma_1}{\gamma}}{\Sym(\gamma) \Sym(\Gamma_1)\Sym(\Gamma_2)} ~ \gamma \Gamma_2 \otimes \Gamma_1 \nn \\ 
& + \hspace{-2mm} \sum_{\gamma,\Gamma_2^\qua\neq \emptyset} \hspace{-2mm} \frac{\ins{\Gamma_2}{\gamma}}{\Sym(\gamma) \Sym(\Gamma_2)} ~\gamma \otimes \Gamma_2
-  \hspace{-3mm} \sum_{\gamma,\Gamma_1^\gluq,\Gamma_2^\qua\neq \emptyset}  \hspace{-2mm} \frac{\ins{\Gamma_2}{\gamma}}{\Sym(\gamma) \Sym(\Gamma_1)\Sym(\Gamma_2)} ~ \gamma \Gamma_1 \otimes \Gamma_2 \nn \\ 
&\qquad - \sum_{\begin{smallmatrix} (\gamma_1,\gamma_2) \neq (\emptyset,\emptyset) \\ \Gamma_1^\gluq,\Gamma_2^\qua \neq \emptyset\end{smallmatrix}} 
\frac{\left(\ins{\Gamma_1}{\gamma}\right) \left(\ins{\Gamma_2}{\gamma}\right)}{\Sym(\gamma_1)\Sym(\gamma_2)\Sym(\Gamma_1)\Sym(\Gamma_2)} ~\gamma_1\gamma_2 \otimes \Gamma_1 \Gamma_2.
\end{align}
The first two terms combine, since
\begin{subequations}
\label{cop:gluq-qua-terms}
\begin{align}
\label{cop:gluq-qua-terms-a}
\sum_\gamma \frac{ \ins{\Gamma_1}{\gamma} }{\Sym(\gamma)} ~\gamma - \sum_{\gamma,\Gamma_2^\qua} \frac{ \ins{\Gamma_1}{\gamma} } {\Sym(\gamma)\Sym(\Gamma_2)} ~ \gamma \Gamma_2 
&= \sum_\gamma \left[ c \begin{pmatrix} \vec{m_{\Gamma_1}} \\ \vec{n_\gamma} \end{pmatrix} - n_{\gamma,\qua} ~ c \begin{pmatrix} \vec{m}_{\Gamma_1} \\ \vec{n}_\gamma- \vec{f}_\qua \end{pmatrix} \right]  \frac{\gamma}{\Sym(\gamma)} \nn \\
&=  \sum_\gamma c \begin{pmatrix} \vec{m}_{\Gamma_1} - \vec{f}_\qua \\ \vec{n}_\gamma \end{pmatrix} \frac{\gamma}{\Sym(\gamma)}~,
\end{align}
by an application of Pascal's rule \eqref{pascal}. On the other hand, the second two terms contribute with
\begin{align}
\label{cop:gluq-qua-terms-b}
\sum_\gamma \frac{ \ins{\Gamma_2}{\gamma} }{\Sym(\gamma)} ~\gamma - \sum_{\gamma,\Gamma_1^\gluq} \frac{ \ins{\Gamma_2}{\gamma} } {\Sym(\gamma)\Sym(\Gamma_1)} ~ \gamma \Gamma_1
&=  \sum_\gamma c \begin{pmatrix} \vec{m}_{\Gamma_2} + \vec{f}_\gluq \\ \vec{n}_\gamma \end{pmatrix} \frac{\gamma}{\Sym(\gamma)}~.
\end{align}
Finally, the quadratic terms becomes 
\begin{align}
\label{cop:gluq-qua-terms-c}
\sum_\gamma \left( \sum_{\vec{n}_1+\vec{n}_2=\vec{n}_\gamma} 
c \begin{pmatrix} \vec{m}_{\Gamma_1} \\ \vec{n}_1 \end{pmatrix}
c \begin{pmatrix} \vec{m}_{\Gamma_2} \\ \vec{n}_2 \end{pmatrix} \right) \frac{\gamma}{\Sym(\gamma)}
= \sum_\gamma
c \begin{pmatrix} \vec{m}_{\Gamma_1}+ \vec{m}_{\Gamma_2} \\ \vec{n}_\gamma \end{pmatrix},
\end{align}
using Vandermonde's identity \eqref{vandermonde}.
\end{subequations}
Unfortunately, the three coefficients in Eq. \eqref{cop:gluq-qua-terms} do not coincide, and it is thus not immediately clear how to combine the terms in Eq. \eqref{cop:gluq-qua} in order to obtain $\ST^\I$ on the second leg of the tensor product. However, we claim that they differ by a finite number of combinations of the steps \I, \II ~and \III, so that they can be combined {\it at the cost of adding terms in} $I \otimes H$, exactly those that appear in Lemma \ref{lma:steps-st}. Indeed, in the case of \eqref{cop:gluq-qua-terms-b} one rewrites trivially, 
\begin{align*}
c \begin{pmatrix} \vec{m}_{\Gamma_2} + \vec{f}_\gluq \\ \vec{n}_\gamma \end{pmatrix} = c \begin{pmatrix} \left( \vec{m}_{\Gamma_2}+ \vec{f}_\gluq +\vec{f}_\qua \right) - \vec{f}_\qua \\ \vec{n}_\gamma \end{pmatrix},
\end{align*}
and observes that the vector $\vec{m}_{\Gamma_2}+ \vec{f}_\gluq +\vec{f}_\qua$ is $(L-K,\gluq)$-admissible if $\gamma$ is at loop order $K$. Also, regarding Eq. \eqref{cop:gluq-qua-terms-c}, one notes that $\vec{m}_{\Gamma_1}+ \vec{m}_{\Gamma_2} + \vec{f}_\qua$ is $(L-K,\gluq)$-admissible. Since $m_{\Gamma_1}$ is $(L-K,\gluq)$-admissible by definition, an application of Lemma \ref{lma:steps} in combination with Lemma \ref{lma:steps-st} shows that all terms in Eq. \eqref{cop:gluq-qua} can be combined to give $h \otimes \ST^\I$ {\it modulo} elements of the form $\sum_A \ST^A \otimes h_A'$ for some $h,h_A' \in H$.

A completely analogous argument shows that the same conclusion holds for $\ST^\II$ and $\ST^\III$.
\end{proof}

\section{Conclusions}
After having derived the necessary combinatorial identities, and obtained a formula for the coproduct on 1PI Green's functions, we have showed that the Slavnov--Taylor identities in quantum chromodynamics can be implemented on the Hopf algebra of Feynman graphs as relations defining a Hopf ideal. Since the map $\phi : H \to K$ defined by the (regularized) Feynman rules (cf. Remark \ref{rem:eff-action}) vanishes on this ideal, it factors through an algebra map from the quotient Hopf algebra $\tilde H:=H/I$ to the field $K$ of Laurent series in the regularization parameter. Since $\tilde H$ is still a commutative (but non-free) connected Hopf algebra, there is a Birkhoff factorization $\phi=\phi_-^{-1} \ast \phi_+$ \cite[Theorem 4]{CK99} for the convolution product $\ast$ in $\Hom(\tilde H,K)$. The two algebra maps $\phi_\pm : \tilde H \to K$ are given on 1PI graphs by the following recursive formula,
\begin{align*}
\phi_-(\Gamma)&=-T \left( \phi(\Gamma)+\sum_{\gamma\subsetneq \Gamma} \phi_-(\gamma) \phi(\Gamma/\gamma) \right),\\
\phi_+(\Gamma)&=\phi(\Gamma) + \phi_-(\Gamma) + \sum_{\gamma\subsetneq \Gamma} \phi_-(\gamma) \phi(\Gamma/\gamma).
\end{align*}
where $T$ is the projection on the pole part in $K$. It was realized in \cite{CK99} that $\phi_+(\Gamma)$ and $\phi_-(\Gamma)$ precisely give the renormalized Feynman amplitude and the counterterms, respectively, corresponding to the graph $\Gamma$. Since they are algebra maps from $\tilde H$ to $K$, we conclude that they automatically satisfy the Slavnov--Taylor identities. 

Moreover, the compatibility of the Slavnov--Taylor identities with the Hopf algebra structure implies validity of the `gauge theory theorem' in \cite[Thm. 5]{Kre05} (see also \cite{Kre06}). The latter states that there exists a certain sub Hopf algebra and that the map $B^\gamma_+: H \to H_\lin, X \mapsto \sum_\Gamma n'(\gamma,X,\Gamma)$, where $n'$ is a normalized version of $n$ defined just above Eq. \eqref{eq:cop-gamma-comb}, is a closed Hochschild cocycle. 

\section{Acknowledgements}
The author would like to thank Matilde Marcolli for several discussions. 

\appendix
\section{An alternative proof of Theorem \ref{thm:ward}}
We give an alternative proof of the above Theorem stating compatibility of the Ward identities with the coproduct, which is based on our previous result \cite{Sui06} on the compatibility of the so-called {\it Ward--Takahashi identities} with the coproduct. Such Ward--Takahashi identities are identities between individual graphs, and an expression of the above Ward elements $W_L$ in terms of the Ward--Takahashi elements (to be introduced below) aloows us to deduce Theorem \ref{thm:ward} from this result. 

Let us start by introducing a map similar to $\Delta_\gamma$ introduced in the Section \ref{sect:ins}. For a disjoint union of 1PI graphs $\gamma$ and a 1PI electron self-energy graph $\gamma_e$, we define on a 1PI graph $\Gamma$,
\begin{equation}
\label{eq:cop-gamma-e}
\Delta_{\gamma,\gamma_e}^e(\Gamma) := \sum_{
\begin{smallmatrix}
\gamma'\gamma'_e \subset \Gamma\\
\gamma'\gamma'_e \isom \gamma\gamma_e
\end{smallmatrix}
}
\Gamma/\gamma'\gamma'_e(e'),
\end{equation}
where $\Gamma/\gamma'\gamma'_e(e')$ is the graph $\Gamma/\gamma'\gamma'_e$ with an external photon line attached to the edge $e'$ corresponding to $\gamma_e'$ in the quotient. In the case that $\gamma$ is the empty graph, this map will be denoted by $\Delta_{\gamma_e}^e := \Delta_{\emptyset,\gamma_e}^e$ and when in addition $\gamma_e = \Gamma$, we set $\Delta_\Gamma^e(\Gamma) =\emptyset$. 
Two examples that might help the reader to see what is going on are, 
\begin{align*}
\Delta^e_{
  \parbox{20pt}{
    \begin{fmfgraph*}(20,11)
      \fmfleft{l}
      \fmfright{r}
      \fmf{plain}{l,v2,v3,r}
      \fmf{photon,left,tension=0}{v2,v3}
  \end{fmfgraph*}}
}
\Bigg(
\parbox{40pt}{
  \begin{fmfgraph*}(40,11)
    \fmfleft{l}
    \fmfright{r}
    \fmf{plain}{l,v1,v2,v3,v4,v5,v6,r}
    \fmf{photon,left,tension=0}{v1,v6}
    \fmf{photon,left,tension=0}{v2,v3}
    \fmf{photon,left,tension=0}{v4,v5}
  \end{fmfgraph*}
}
\Bigg)&=~
\parbox{40pt}{\begin{fmfgraph*}(40,20)
    \fmfforce{(.33w,0h)}{b}
    \fmfleft{l}
    \fmfright{r}
    \fmf{plain}{l,v1,v3,v5,v6,v7,r}
    \fmffreeze
    \fmf{photon}{b,v3}
    \fmf{photon,left,tension=0}{v5,v6}
    \fmf{photon,left,tension=0}{v1,v7}
    \end{fmfgraph*}
}
~+~
\parbox{40pt}{\begin{fmfgraph*}(40,20)
    \fmfforce{(.66w,0h)}{b}
    \fmfleft{l}
    \fmfright{r}
    \fmf{plain}{l,v1,v2,v3,v6,v7,r}
    \fmffreeze
    \fmf{photon}{b,v6}
    \fmf{photon,left,tension=0}{v2,v3}
    \fmf{photon,left,tension=0}{v1,v7}
    \end{fmfgraph*}
}~, \\
\Delta^e_{
  \parbox{20pt}{
    \begin{fmfgraph*}(20,11)
      \fmfleft{l}
      \fmfright{r}
      \fmfbottom{b}
      \fmf{plain}{l,v1,v2,v3,r}
      \fmf{photon,left,tension=0}{v1,v3}
      \fmffreeze
      \fmf{photon}{b,v2}
  \end{fmfgraph*}}
~,~  \parbox{20pt}{
    \begin{fmfgraph*}(20,11)
      \fmfleft{l}
      \fmfright{r}
      \fmf{plain}{l,v2,v3,r}
      \fmf{photon,left,tension=0}{v2,v3}
  \end{fmfgraph*}}
}
\Bigg(
\parbox{40pt}{
  \begin{fmfgraph*}(40,11)
    \fmfleft{l}
    \fmfright{r}
    \fmf{plain}{l,v1,v2,v3,v4,v5,v6,r}
    \fmf{photon,left,tension=0}{v1,v5}
    \fmf{photon,left,tension=0}{v2,v3}
    \fmf{photon,right,tension=0}{v4,v6}
  \end{fmfgraph*}
}
\Bigg)&=~ \parbox{40pt}{
    \begin{fmfgraph*}(40,11)
      \fmfleft{l}
      \fmfright{r}
      \fmfbottom{b}
      \fmf{plain}{l,v1,v2,v3,r}
      \fmf{photon,left,tension=0}{v1,v3}
      \fmffreeze
      \fmf{photon}{b,v2}
  \end{fmfgraph*}}~.
\end{align*}
The analogue of $n(\Gamma,\gamma,\Gamma')$ is given by $n^e(\Gamma, \gamma, \Gamma')$; it is defined to be the number of subgraphs $\gamma'$ of the electron self-energy graph $\Gamma$ that are isomorphic to the electron self-energy graph $\gamma$, and such that $\Gamma/\gamma'(e')$ is isomorphic to the vertex graph $\Gamma'$. 

We also introduce the following combinatorial factor, to be used below. 
The compatibility of the so-called Ward--Takahashi identities with the coproduct was derived in \cite{Sui06} and will be recalled in Proposition \ref{prop:wt} below; for completeness, we also restate its proof. It gives relations between individual graphs, and will be used in the alternative proof of Theorem \ref{thm:ward}, involving relations between Green's functions. We first recall some of the notation in that paper. 

Given any electron self-energy graph $\Gamma$, we can label the internal electron edges from $1$ to $m_{\Gamma,\el}$. If we fix such a labelling, we denote by $\Gamma(i)$ the graph $\Gamma$ with an external photon line attached to the electron line $i$. 

\begin{prop}
\label{prop:wt}
Let $\Gamma^\el$ be a 1PI electron self-energy graphs and define the corresponding {\rm Ward--Takahashi element} by $W(\Gamma)=\sum_i \Gamma(i) + \Gamma$, with the sum over internal electron lines in $\Gamma$, Moreover, set $W(\emptyset)=0$. Then,
\begin{equation*}
\Delta\left(W(\Gamma)\right) = \sum_{\gamma,\gamma_e^\el} \frac{1}{|\gamma\gamma_e|_\vee}  ~ W(\gamma_e) \gamma \otimes \Delta^e_{\gamma,\gamma_e}(\Gamma)
+\sum_\gamma \frac{1}{|\gamma|_\vee}  ~ \gamma \otimes W(\Delta_\gamma(\Gamma)),
\end{equation*}
where the sum is over disjoint unions of 1PI graphs $\gamma$ (including the empty graph) and 1PI electron self-energy graphs $\gamma_e$. 
\end{prop}
\begin{proof}
We start by computing $\Delta\left(\sum_i \Gamma(i) \right)$; we split the sum over subgraphs into two terms, those for which $\gamma$ contains the electron line $i$, and those for which it does not,
\begin{align*}
\Delta\left(\sum_i \Gamma(i) \right) \equiv \sum_{i=1}^{m_{\Gamma,\el}} \sum_{\gamma \subseteq \Gamma(i)} \gamma \otimes \Gamma(i)/\gamma 
=\sum_{\gamma \subseteq \Gamma} \left( \sum_{i \in \gamma_E} \gamma(i) \otimes \Gamma(i)/\gamma(i) + \sum_{i \notin \gamma} \gamma \otimes \Gamma(i) /\gamma \right),
\end{align*}
where $\gamma_E$ denotes the disjoint union of the electron self-energy graphs in $\gamma$. The absence of terms in which $i \in \gamma-\gamma_E$ is due to the fact that there are no vertices with more than one photon line, so that the corresponding graph $(\gamma-\gamma_E)(i)$ does not appear as a subgraph of $\Gamma(i)$ in the coproduct. 

The first term can be written as,
\begin{align*}
\sum_{\gamma \subseteq \Gamma}  \sum_{i \in \gamma_E} \gamma(i) \otimes \Gamma(i)/\gamma(i)
&=\sum_{\gamma \subseteq \Gamma} \sum_{\gamma_e \subset \gamma} (\gamma-\gamma_e) \sum_{i\in \gamma_e}\gamma_e(i) \otimes \Gamma/\gamma\gamma_e (e),
\end{align*}
by decomposing the sum over all $i$ that are part of the subgraph $\gamma_E$ into the connected components $\gamma_e$ of $\gamma_E$, thereby also noting that if $i \in \gamma_e$, the quotient $\Gamma(i)/\gamma_e(i)$ is nothing else then $\Gamma/\gamma_e(e)$. 

On the other hand, the second term can be split into two parts, one for which $i$ is an external electron edge for one of the electron self-energy graphs in $\gamma$, and one for which it is not,
\begin{align*}
\sum_{\gamma \subseteq \Gamma} \sum_{i \notin \gamma} \gamma \otimes \Gamma(i) /\gamma
&=\sum_{\gamma \subseteq \Gamma} \left( \sum_{i \in \partial \gamma_E} \gamma \otimes \Gamma(i)/\gamma + \sum_{i \notin \gamma \cup \partial \gamma_E} \gamma \otimes \Gamma/\gamma(i)   \right) \\
&= \sum_{\gamma \subseteq \Gamma} \sum_{\gamma_e \subset \gamma}  (\gamma-\gamma_e) \gamma_e \otimes \Gamma/(\gamma-\gamma_e)\gamma_e(e) +  \gamma \otimes \sum_{j\in\Gamma/\gamma} \Gamma/\gamma(j).
\end{align*}
Here we used the fact that although an electron self-energy graph has two external electron edges, they might be in common with another electron self-energy graph that is also part of $\gamma$. This means that for a block of electron self-energy graphs that sit inside $\Gamma$ by concatenation, there is a term $(\gamma-\gamma_e) \gamma_e \otimes \Gamma/(\gamma-\gamma_e)\gamma_e(e)$ for each of such graphs, plus one for the closing ``fence pole''. Indeed, the number of external electron edges of such a block is precisely the number of its constituents plus one. The extra term combines with the second term, since it precisely gives the term $\Gamma/\gamma(j)$ with $j$ the electron edge corresponding to $\gamma_e$ so as to complete the sum over all internal electron edges in the last term. 
\end{proof}

\begin{prop}
\label{eq:W-WT}
There are the following relations between $W(\Gamma)$ and the Ward elements $W_L$,
\begin{equation*}
W_L = \sum_{\Gamma_L^\el}\frac{1}{\Sym(\Gamma)} ~  W(\Gamma).
\end{equation*}
\end{prop}
\begin{proof}
Since every vertex graph can be written as $\Gamma(i)$ for an electron self-energy graph $\Gamma$, we only have to check that the symmetry factors turn out right. We first observe that $\Sym(\Gamma(i))$ coincides with the order of the isotropy group $\Aut(\Gamma)_i$ of the electron edge $i$ in $\Aut(\Gamma)$. Moreover, two graphs $\Gamma(i)$ and $\Gamma(i')$ are isomorphic if and only if $i$ and $i'$ are in the same orbit under the action of $\Aut(\Gamma)$. An application of the orbit-stabilizer theorem shows that such an orbit has length $|\Aut(\Gamma)i|=\Sym(\Gamma) / |\Aut(\Gamma)_i | 
$, so that,
\begin{equation*}
\sum_{\Gamma_L^\el} \sum_i \frac{1}{\Sym(\Gamma)} ~ \Gamma(i) = \sum_{\Gamma_L^\el} \sum_i \frac{1}{|\Aut(\Gamma)i| \Sym(\Gamma(i))} ~ \Gamma(i) = \sum_{\Gamma_L^\vertex} \frac{1}{\Sym(\Gamma)} ~ \Gamma,
\end{equation*}
from which the result follows at once.
\end{proof}
Before continuing our derivation of Theorem \ref{thm:ward} from Proposition \ref{prop:wt}, we derive a result anologous to Lemma \ref{lma:defect} for the map $\Delta_{\gamma,\gamma_e}^e$. Instead of the insertion of graphs of Definition \ref{defn:ins} that appeared in Lemma \ref{lma:defect}, it involves insertion of an electron self-energy graph into a vertex graph, defined as follows.
\begin{defn}
\label{defn:insertion}
An insertion of a 1PI electron self-energy graph $\gamma$ into a 1PI vertex graph $\Gamma$ is given by an isomorphism $\phi$ from $\gamma^\e_\exter$ to the two electron lines connected to the vertex to which the external photon line is attached, after removal of this photon line. The resulting electron self-energy graph is denoted by $\Gamma \cross{\phi} \gamma$. 
\end{defn}
\noindent For example,
$$
\parbox{50pt}{\begin{fmfgraph*}(50,20)
    \fmfbottom{b}
    \fmfleft{l}
    \fmfright{r}
    \fmf{plain}{l,v1,v2,v3,r}
    \fmffreeze
    \fmf{photon}{b,v2}
    \fmf{photon,left,tension=0}{v1,v3}
    \end{fmfgraph*}}
\cross{\phi}
\parbox{50pt}{\begin{fmfgraph*}(50,20)
    \fmfleft{l}
    \fmfright{r}
    \fmf{plain}{l,v1,v3,r}
    \fmffreeze
    \fmf{photon,left,tension=0}{v1,v3}
    \end{fmfgraph*}}
~=~ \parbox{50pt}{\begin{fmfgraph*}(50,20)
    \fmfleft{l}
    \fmfright{r}
    \fmf{plain}{l,v1,v2,v3,v4,r}
    \fmffreeze
    \fmf{photon,left,tension=0}{v1,v4}
    \fmf{photon,left,tension=0}{v2,v3}
    \end{fmfgraph*}}~.
$$

\begin{lma}
\label{lma:defect-e}
Let $\gamma$ and $\gamma_e$ as above and let $n_{\gamma,\vertex}$ denote the number of 1PI vertex graphs in $\gamma$ as before. Then,
\begin{equation*}
\Delta^e_{\gamma,\gamma_e} = \Delta_\gamma \Delta^e_{\gamma_e} - \tilde, \rho_{\gamma,\gamma_e}
\end{equation*}
where $\tilde \rho_{\gamma,\gamma_e}$ is defined in terms of the vertex graph $\gamma_1, \ldots, \gamma_{n_{\gamma,\vertex}}$ in $\gamma$ by,
\begin{equation*}
\tilde \rho_{\gamma,\gamma_e}=\sum_{l=1}^{n_{\gamma,\vertex}} \sum_{[\phi]}\frac{n^e(\gamma_l \cross{\phi} \gamma_e,\gamma_e,\gamma_l)}{n(\gamma-\gamma_l,\gamma_l)} \frac{|\gamma_l \cross{\phi} \gamma_e|_\vee}{|\gamma_l|_\vee} ~ \Delta^e_{\gamma-\gamma_l, \gamma_l \cross{\phi} \gamma_e}.
\end{equation*}
\end{lma}
\begin{proof}
Similar to the proof of Lemma \ref{lma:defect}, we introduce a map $\tilde\rho_{\gamma,\gamma_e}$; it 
corrects for the overcounting in the cases that there is a subgraph $\tilde\gamma$ of $\Gamma$ containing $\gamma'_e\isom \gamma_e$, such that $\tilde\gamma/\gamma'_e(e') \isom \gamma$. Indeed, quotienting $\Gamma$ by such a $\tilde\gamma$ does not appear in $\Delta_{\gamma,\gamma_e}^e(\Gamma)$, although it does in $\Delta_\gamma \Delta^e_{\gamma_e}(\Gamma)$. We then write, 
\begin{align*}
\sum_{\begin{smallmatrix}\gamma'_e \subset \tilde\gamma\subset\Gamma \\ \tilde\gamma/\gamma'_e (e') \isom \gamma \end{smallmatrix}} \Gamma/\tilde\gamma
= \sum_{l=1}^{n_{\gamma,\vertex}} \sum_{[\phi]} \frac{n^e(\gamma_l \cross{\phi} \gamma_e,\gamma_e,\gamma_l)}{n(\gamma-\gamma_l,\gamma_l)} \frac{|\gamma_l \cross{\phi} \gamma_e|_\vee}{|\gamma_l|_\vee} ~\Delta^e_{\gamma-\gamma_l,\gamma_l \cross{\phi} \gamma_e}(\Gamma),
\end{align*}
since each such graph $\tilde\gamma$ has to be isomorphic to the graph obtained by inserting $\gamma_e$ (using $\#$) into one of the vertex graph components $\gamma_l^\vertex$ of $\gamma$, with an additional factor of $n^e(\gamma_l \cross{\phi} \gamma_e,\gamma_e,\gamma_l)$ due to the multiplicity arising on the left-hand-side. On the other hand, division by $n(\gamma-\gamma_l,\gamma_l)$ corrects for insertions of $\gamma_e$ into isomorphic $\gamma_l$; indeed, if $\gamma_l \isom \gamma_{l'}$ then clearly $\gamma_l \cross{\phi} \gamma_e \isom \gamma_{l'} \cross{\phi} \gamma_e$. Finally, the factor $|\gamma_l \cross{\phi} \gamma_e|_\vee/|\gamma_l|_\vee$ arises from the difference of summing over subgraphs isomorphic to $\gamma_l$ and subgraphs isomorphic to $\gamma_l \cross{\phi} \gamma_e$.
\end{proof}

\begin{thm-ward}
For any $L \geq 0$, we have 
\begin{multline*}
\Delta(W_L) =  \hspace{-2mm}  \sum_{K+K'=0}^N  \hspace{-2mm} W_K \sum_{\gamma_{K'}} \frac{c(L-K-K',\gamma)}{\Sym(\gamma)} ~ \gamma \otimes G^\vertex_{L-K-K'} 
+ \sum_{K=0}^L \sum_{\gamma_{K}} \frac{\ins{(L-K,\el)}{\gamma}}{\Sym(\gamma)}~ \gamma \otimes W_{L-K}.
\end{multline*}
Consequently, the ideal $I$ generated by the Ward elements $W_L$ for every $L$ is a Hopf ideal in $H$,
\begin{gather*}
\Delta(I) \subseteq I \otimes H + H \otimes I ,\qquad \epsilon(I)=0,\qquad S(I) \subseteq I.
\end{gather*}
\end{thm-ward}
\begin{proof}
Firstly, with $W_L$ being related to $W(\Gamma)$ by the above Eq. \eqref{eq:W-WT} (and understood to be zero for $L=0$), Proposition \ref{prop:wt} implies that 
\begin{equation*}
\Delta(W_L)
= \sum_{K+K'=0}^L 
\sum_{\begin{smallmatrix}\Gamma_{L}^\el \\ \gamma_{K'},\gamma^\el_{e,K} \end{smallmatrix}} \frac{1}{|\gamma\gamma_e|_\vee\Sym(\Gamma)} ~  W(\gamma_e) \gamma \otimes \Delta^e_{\gamma,\gamma_e}(\Gamma)
+(\id \otimes W ) \left( \sum_{\Gamma_L^\el} \frac{1}{\Sym(\Gamma)} ~ \Delta(\Gamma) \right).
\end{equation*}
The second term can be easily written in the desired form since with Proposition \ref{prop:linear},
\begin{align*}
(\id \otimes W ) \left( \sum_{\Gamma_L^\el} \frac{1}{\Sym(\Gamma)} ~ \Delta(\Gamma) \right) 
&= \sum_{K=0}^L \sum_{\gamma_K,\Gamma_{L-K}^\el} \frac{\ins{\Gamma}{\gamma}}{\Sym(\gamma)\Sym(\Gamma)}  ~ \gamma \otimes W(\Gamma)\\
&= \sum_{K=0}^L\sum_{\gamma_K} \frac{\ins{(L-K,\el)}{\gamma}}{\Sym(\gamma)}  ~ \gamma \otimes W_{L-K},
\end{align*}
which is indeed an element in $H \otimes I$. For the first term, we derive the result from the following equation,
\begin{equation}
\label{eq:ind}
\sum_{\Gamma_L^\el} \frac{1}{|\gamma\gamma_e|_\vee\Sym(\Gamma)} ~ \Delta^e_{\gamma,\gamma_e}(\Gamma) =\sum_{\Gamma_{L-K-K'}^\vertex} \frac{c(L-K-K',\gamma)}{\Sym(\gamma_e)\Sym(\gamma)\Sym(\Gamma)} ~ \Gamma,
\end{equation}
for two graphs $\gamma$ and $\gamma_e$ at respective loop order $K'$ and $K$. In fact, validity of this equation implies that
\begin{multline*}
\sum_{K,K'} 
\sum_{\begin{smallmatrix}\Gamma_{L}^\el \\ \gamma_{K'},\gamma^\el_{e,K}\end{smallmatrix}} 
\frac{1}{|\gamma\gamma_e|_\vee\Sym(\Gamma)} ~ W(\gamma_e) \gamma \otimes \Delta^e_{\gamma,\gamma_e}(\Gamma)= \sum_{K,K'} W_K \!\!\!\! \sum_{\gamma_{K'},\Gamma_{L-K-K'}^\vertex} \!\!\!\!\frac{c(L-K-K',\gamma) }{\Sym(\gamma)\Sym(\Gamma)} ~ \gamma \otimes \Gamma,
\end{multline*}
which, on its turn, is an element in $I \otimes H$. We prove that equation \eqref{eq:ind} holds by induction on the number $n_{\gamma,\vertex}$ of 1PI vertex graphs in $\gamma$. If this number is zero, Lemma \ref{lma:defect-e} takes a simple form and we conclude that in this case,
\begin{equation}
\label{eq:ind0}
\sum_{\Gamma_L^\el} \frac{1}{|\gamma\gamma_e|_\vee\Sym(\Gamma)} ~ \Delta^e_{\gamma,\gamma_e}(\Gamma)
= \sum_{\tilde\Gamma_{L-K}^\vertex} \frac{1}{|\gamma|_\vee \Sym(\gamma_e) \Sym(\tilde\Gamma)} ~ \Delta_{\gamma}(\tilde\Gamma),
\end{equation}
since for each 1PI vertex graph $\tilde\Gamma$, there are precisely $|\gamma_e|_\vee |\Aut(\Gamma)_i|=|\gamma_e|_\vee \Sym(\Gamma(i))$ many electron self-energy graphs $\Gamma$ that result in $\tilde\Gamma$ after quotienting by a subgraph isomorphic to $\gamma_e$ and connecting an external photon line to the corresponding internal electron line. Thus, an application of Equation \eqref{eq:cop-gamma} shows validity of Eq. \eqref{eq:ind}, since by definition $c(L-K-K',\gamma)=\ins{(L-K-K',\vertex)}{ \gamma}$. 

Let us then suppose that Eq. \eqref{eq:ind} holds for $n_{\gamma,\vertex}=n$ and number the vertex graphs in $\gamma$ by $\gamma_1, \ldots, \gamma_n$. If $\gamma_{n+1}$ is another 1PI vertex graph, then Lemma \ref{lma:defect-e} yields
\begin{multline*}
\sum_{\Gamma_L^\el} \frac{1}{|\gamma\gamma_{n+1}\gamma_e|_\vee\Sym(\Gamma)} ~ \Delta^e_{\gamma\gamma_{n+1},\gamma_e}(\Gamma) 
=  \sum_{\Gamma_L^\el} \frac{1}{|\gamma\gamma_{n+1}\gamma_e|_\vee\Sym(\Gamma)} ~ \Delta_{\gamma\gamma_{n+1}}\Delta^e_{\gamma_e}(\Gamma) \\
- \sum_{\Gamma_L^\el} \sum_{l=1}^{n+1} \sum_{[\phi]} \frac{n^e(\gamma_l \cross{\phi} \gamma_e,\gamma_e,\gamma_l) }{|\gamma\gamma_{n+1}-\gamma_l|_\vee |\gamma_l \cross{\phi} \gamma_e|_\vee \gamma_e|_\vee \Sym(\Gamma) n(\gamma\gamma_{n+1}-\gamma_l,\gamma_l)} \sum_{[\phi]} \Delta^e_{\gamma\gamma_{n+1}-\gamma_l, \gamma_l \cross{\phi} \gamma_e} (\Gamma),
\end{multline*}
after insertion of the definition of $\tilde\rho$ and an application of Definition \ref{defn:triv-perm}. The first term can be reduced by applying Eq. \eqref{eq:ind0} to $\Delta^e_{\gamma_e}$ and Eq. \eqref{eq:cop-gamma} to $\Delta_{\gamma\gamma_{n+1}}$. For the second term, we can apply the induction hypothesis, since $\gamma\gamma_{n+1}-\gamma_l$ has $n$ 1PI vertex graph components. With the following two equalities for the combinatorial factors:
\begin{gather*}
n(\gamma \gamma_{n+1}-\gamma_l, \gamma_l )\Sym(\gamma_l)\Sym(\gamma \gamma_{n+1} - \gamma_l)=\Sym(\gamma \gamma_{n+1});\\
\Sym(\gamma_l \cross{\phi} \gamma_e) = \Sym(\gamma_e) \Sym(\gamma_l) ~ n^e(\gamma_l \cross{\phi} \gamma_e,\gamma_e,\gamma_l),
\end{gather*}
which can be obtained from Eq. \eqref{eq:sym-union} and the observation that $n^e(\gamma \cross{\phi} \gamma_e,\gamma_e, \gamma)$ counts precisely the automorphisms of $\gamma_l \cross{\phi}\gamma_e$ that do not come from an automorphism of $\gamma_l$ or $\gamma_e$, we derive
\begin{equation*}
\sum_{\Gamma_L^\el} \frac{1}{|\gamma\gamma_{n+1}\gamma_e|_\vee\Sym(\Gamma)} ~ \Delta^e_{\gamma\gamma_{n+1},\gamma_e}(\Gamma) = \sum_{\Gamma_{L-K-K'}^\vertex} \frac{\ins{\Gamma}{\gamma\gamma_{n+1}} -  \sum_{l=1}^{n+1} c(L-K-K',\gamma \gamma_{n+1} - \gamma_l) }{\Sym(\gamma_e)\Sym(\gamma\gamma_{n+1}) \Sym(\Gamma)} ~ \Gamma.
\end{equation*}
Note that the sum over $[\phi]$ has been cancelled against the factor $|\gamma_e|_\vee$ in the denominator. A glance back at the recursive definition of $c$ in Equation \eqref{eq:c-rec} then completes the proof. 
\end{proof}

\section{Some combinatorial identities}
For completeness, we list some combinatorial identities used throughout the text. Firstly, there is the well-known Pascal's rule:
\begin{subequations}
\label{pascal}
\begin{align}
{ k+1 \choose l } = { k \choose l } + { k \choose l-1 },
\end{align}
which can be conveniently rewritten as
\begin{align}
{ k+l-2 \choose l } = { k+l-1 \choose l } - { k+l-2 \choose l-1 },
\end{align}
\end{subequations}
for combinations with repetition. 

Another identity used in the text is due to Vandermonde, stating that
\begin{subequations}
\label{vandermonde}
\begin{align}
\sum_{l=0}^q {k_1 \choose l} {k_2 \choose q-l} = {k_1 + k_2 \choose q},
\end{align}
which can be proved by expanding both sides of $(1+t)^{k_1+k_2} = (1+t)^{k_1} (1+t)^{k_2}$. 
The analogous result for combinations with repetition is,
\begin{align}
\sum_{l=0}^q {k_1+l-1 \choose l} {k_2+q-l-1 \choose q-l} = {k_1 + k_2+q-1 \choose q}.
\end{align}
\end{subequations}
This can be proved by equating both sides of $(1-t)^{-k_1-k_2}=(1-t)^{-k_1} (1-t)^{-k_2}$, where $(1-t)^{-n} = \sum_r { n+r-1 \choose r } t^r$ is the generating function for combinations with repetition. 

\end{fmffile}


\end{document}